\documentstyle[10pt]{article}
\input{psfig}
\def\half{{\frac{1}{2}}}

\newcommand{\keiko}{\stackrel{\triangle}{=}}

\newcommand{\implies}{\ \Rightarrow\ }

\newcommand{\xhat}{\hat{x}}

\newcommand{\ie}{{\em i.e.}}
\newcommand{\eg}{{\em e.g.}}
\newcommand{\etc}{{\em etc.}}

\newcounter{thanksnum}
\def\thanksnumber#1
{\setcounter{thanksnum}{\value{footnote}}\setcounter{footnote}{#1}%
\addtocounter{footnote}{-1}\footnotemark
\setcounter{footnote}{\value{thanksnum}}}
 
\newcommand{\yhat}{\hat{y}}

\newcommand{\uhat}{\hat{u}}

\begin{document}
\begin{center}

{\Large\bf A Survey of Manoeuvring Target Tracking Methods
\footnote{Original manuscript dated 23/12/1998. Submitted to IEEE Transactions on Aerospace \& Electronic Systems. Withdrawn following a review process lasting 3 years after it became clear that there was a conflict of interest on the part of the Associate Editor handling the submission.}
}
\vskip 12mm
{\large\bf Graham W. Pulford$^\dagger$}\\
\vskip 8mm
{\baselineskip=18pt $^\dagger$Department of Electrical and Electronic
Engineering, University of Melbourne, Parkville 3052, Australia and
the Cooperative Research Centre for Sensor Signal and Information
Processing (CSSIP)}
\vskip 5mm
\today
\vskip 5mm

\end{center}

\date{22 December 1998}

\tableofcontents

 \newpage
\begin{center}
{\Large\bf Abstract}
\end{center}

A comprehensive review of the literature on manoeuvring target
tracking for both uncluttered and cluttered measurements is presented.
Various discrete-time dynamical models including non-random input, random-input and
switching or hybrid system manoeuvre models are presented. The problem
of manoeuvre detection is covered. Classical and current filtering
methods for manoeuvre tracking are presented including multi-level
process noise, input estimation, variable dimension filtering,
two-stage Kalman filter, the interacting multiple model algorithm, and
generalised pseudo-Bayesian algorithms. Various extensions of these
algorithms to the case of cluttered measurements are also described
and these include: joint manoeuvre and measurement association,
probabilistic data association and multi-hypothesis tracking. Smoothing
schemes, including IMM smoothing and batch expectation-maximisation
using the Viterbi algorithm, are also described. The use of amplitude
information for target measurement discrimination is discussed. It is
noted that although many manoeuvre tacking techniques exist, the
literature contains surprisingly few performance comparisons to guide
the design engineer although a performance benchmark has recently been
introduced.

\section{Introduction}\label{intro}
The problem of tracking a manoeuvring target based on noisy sensor
measurements has been approached by a great many researchers in the
last few decades.  The comprehensive taxonomy of the existing
approaches presented in this article should convince the reader that
this is a very active research area. Manoeuvres correspond to unknown
acceleration terms in the target dynamics, increasing the number of
quantities that must be estimated compared with the constant velocity
case.  The fundamental problems in manoeuvring target tracking are (i)
manoeuvre detection: the detection of the onset (and end) of a
manoeuvre, and (ii) compensation: the correction of target state
estimates to allow for the manoeuvre. A third problem of tracking a
manoeuvring target in clutter is discussed subsequently.  Classical
techniques based on adjustments or slight modifications to the
discrete-time Kalman filter include increasing the process noise
covariance during a manoeuvre, finite memory (sliding data window) and
fading memory (exponential forgetting) filters \cite{Jazwinski}. All
of these approaches increase the effective filter bandwidth so as to
respond more rapidly to the measurement data during a manoeuvre; but
they are risky in the presence of clutter since they increase the size
of the validation gate.

Some of the more recent approaches to manoeuvring target tracking include:
adaptive Kalman filters \cite{Hampton,Chang5},
filters using correlated and semi-Markov process noise
\cite{Singer,Singer1,Guu1} usually implemented using
 multiple model (partitioning) filters and filter banks
\cite{Thorp,Gholson,Moose2,Ricker,SunChiang,WangHP}; 
filters based on Poisson and renewal process models of acceleration
\cite{Lim,Sworder2}; 
input estimation \cite{Chan,Chan1} and input and onset-time estimation
\cite{Bogler,Chang3,Sastry}; variable dimension filters
\cite{Barshalom11}; track splitting filters with a finite
memory constraint
\cite{West}; the generalised pseudo-Bayesian (GPB) algorithm
\cite{Jaffer,Ackerson}; and the interacting multiple model
(IMM) algorithm \cite{Blom,Blair1,Munir1}.  A second-order extension
of the IMM algorithm was developed in \cite{Blair4}.  

Some researchers have developed approaches to avoid the two-stage nature
of manoeuvre detection and compensation implied by methods like
variable dimension filtering. These include concurrent input
estimation (of acceleration and manoeuvre onset time) and
variable-dimension filtering \cite{Park}; and an interacting multiple
model filter including a bias estimator
\cite{Watson3,Blair3,Alouani} which treats the acceleration as
a bias term to be estimated and used for compensation of the target
state estimate. A filter developed in \cite{Alouani1} uses a kinematic
speed constraint in the measurement equation for tracking constant-speed
manoeuvring targets (the acceleration should be orthogonal to the
velocity).

A brief overview of some of these techniques is contained in
\cite{Farooq2}, while a detailed treatment of the filter bank approach
using a semi-Markov process noise model may be found in \cite{Farina}.
A comparison of input/onset-time estimation with the interacting
multiple model algorithm appeared in \cite{Barshalom11} (with
follow-up comments contained in \cite{Farooq3} and \cite{Hou}),
demonstrating the comparable performance of the IMM algorithm at lower
computational complexity.  Several methods based on partitioning
(multiple model) filters have been reviewed in the context of
bearings-only tracking for non-manoeuvring and manoeuvring targets in
\cite{Katsikas}, and their performance compared with that of a single
Kalman filter technique.

A technique for complexity reduction of existing manoeuvring target
trackers based on the information form of the Kalman filter was
presented in \cite{Rouhi,Farooq1}. Manoeuvres were modelled as
continuous-time stochastic process with known prior density and a
conditional mean estimator derived in \cite{Benes}; the estimator is
implementable when the input process has a finite number of discrete
values with known prior probabilities.

A comparison of Kalman and $H_{\infty}$ filters for
tracking constant-speed manoeuvring targets recently appeared in
\cite{Tsaknakis}.  The conclusion reached was that an ``untuned''
$H_{\infty}$ filter does not outperform an appropriately tuned Kalman
filter, and that the $H_{\infty}$ filter tends to be more robust to mistuning. 
The issue of tuning in $H_{\infty}$ filtering is important firstly
in that it requires the selection of an extra parameter $\gamma$, and
secondly since the filter is not realisable if $\gamma$ is made too small.

The literature on manoeuvring target tracking is not restricted to
filtering as evidenced by work on fixed-interval smoothing using
interacting multiple models \cite{Helmick}; and measurement concatenation
(block processing) applied to variable dimension filtering
\cite{Cloutier1,Cloutier2,Cloutier}.  As remarked in \cite{Farooq2},
the variable dimension filtering approach is in fact a smoothing
technique, since it requires the accelerating target model to be
initialised retrospectively, \ie, using measurements prior
to the instant of manoeuvre detection.

There are also off-line approaches to the problem of manoeuvring
target tracking which require a batch of measurement data for
processing.  The bearings-only tracking problem for a manoeuvring
target has been attempted via a batch method that uses simulated
annealing to initialise an optimisation routine that determines
maximum likelihood state estimates. The Expectation-Maximisation (EM)
algorithm \cite{Baum2,Titterington,Tanner} has recently been
applied by the author to determine the maximum {\em a posteriori}
sequence of target manoeuvres in a multi-level white noise model
\cite{Pulford10}. The EM algorithm is a multi-pass technique
that requires a complete track estimate for initialisation.  A
recursive, sub-optimal version of the preceding algorithm has also
been developed and has been shown to be closely related to the Viterbi
manoeuvre tracker in \cite{Averbuch}.

Some less conventional approaches to manoeuvring target tracking
include fuzzy systems and artificial neural networks.  The fuzzy
logicians \cite{Kim,Ott} have concentrated on emulating alpha-beta
trackers, which are simplified Kalman filters using precomputed
steady-state gains.  Approaches based on artificial neural
networks have considered estimation or modelling of the target
acceleration using neural nets \cite{Amoozegar}, or replacing the
entire Kalman filter or PDA algorithm with a neural network
\cite{Tao,Zhongliang,Zhongliang1}. The simulations reported in
these articles are unconvincing or lack any comparison with accepted
techniques. It is also unclear how the important issue of tuning or
``learning'' should be dealt with, since algorithms based on neural
networks often have many adjustable parameters.

\vskip 2mm

The considerably more difficult problem of tracking a manoeuvring
target in a cluttered environment has also received much attention.
As well as unknown acceleration terms with unknown onset times, the
tracker must solve the additional problem of data association, that
is, determining which of several possible measurements is to be
associated with the target.
In the absence of target manoeuvres, tracking in clutter is often
achievable using extensions of Kalman filtering such as the
probabilistic data association (PDA) filter
\cite{Barshalom1} or other, more computationally intensive techniques
\cite{Barshalom2}. These algorithms must be modified to maintain
track during target manoeuvres.
It is clear that the objectives of manoeuvre detection and
data association are somewhat in conflict, since the appearance of a
false measurement can easily lead to an incorrect assumption of a
target manoeuvre.  
The classical approach of nearest-neighbour Kalman filtering
often suffers from filter divergence in cluttered environments.
Recent PDA-type approaches which have been applied to
tracking manoeuvring targets in clutter include:
Bayesian adaptive filters with multi-level white or coloured noise
\cite{Gauvrit,Tomasini,Tomasini1,Tomasini2}; joint measurement and
manoeuvre association filters for a single target \cite{LeeHJ}
and for multiple targets \cite{Sengupta,Kosuge}; variable
dimension Kalman and PDA filters \cite{Birmiwal,Slocumb}; interactive
multiple model PDA \cite{Barshalom4,Barshalom5} and IMM integrated PDA
\cite{Helmick1}; PDA with ``prediction-oriented'' multiple models
\cite{Musicki3}; PDA filters using ranking information
\cite{Nagarajan} or amplitude information \cite{Lerro1,Lerro1a} to
enhance the discrimination of targets from clutter.  Limited-memory
(N-scan-back) 
filters featuring multi-level white noise with Markov switching are
developed in \cite{Kenefic,Walton}. 

A good review of manoeuvring target tracking techniques, together with
some comparative studies of algorithm performance is contained in
\cite{Barshalom2,Barshalom7}.  A MHT tracker using an adaptive
re-initialisation procedure is described in \cite{Quach1}; this
approach has been applied in target motion analysis of acoustic
sources in clutter and performs better than the IMM at low SNR's. A
JPDA approach to multiple, manoeuvring target tracking using an
innovations-based manoeuvre detector and compensator was presented in
\cite{ChungY}; however the simulations assumed an unrealistically low
clutter density.

Trackers based on hidden Markov models, using
the Viterbi algorithm \cite{Forney2} to perform an efficient search over a discretised
state space, have been used to track targets with limited
manoeuvrability in the presence of false alarms and other interference
\cite{Martinerie1,Demirbas}. In another paper \cite{Demirbas1}
a HMM-based tracker is presented, but the emphasis is on the ability
of the technique to handle non-linear measurement models
rather than on manoeuvring in clutter.
The discrete state-space HMM trackers developed in
\cite{Streit2,Xie1,Xie2} for single and multiple frequency line
tracking, which work for time-varying tones, are known to be
adaptable to the cluttered measurement case by suitable modification
of the observation probability density, although this is not
explicitly stated therein. Barniv's dynamic programming algorithm
\cite{Barniv1,Barniv2} for tracking low SNR targets in infra-red image
sequences was extended in \cite{Arnold} to allow for cluttered measurements. 

The Viterbi algorithm can also be applied to tracking problems {\em without}
discretisation of the state and measurement spaces.  A first approach
along these lines is contained in \cite{Quach} in the context of sonar
tracking.  A subsequent study, reported in
\cite{Pulford7,Pulford8,Lascala6}, extends the previous work
to include automatic track maintenance and measurement
gating. This technique, called {\em Viterbi data association} (VDA)
has been applied successfully to the data association problem in
over-the-horizon radar (OTHR) and more recently extended to allow for
manoeuvring targets \cite{Lascala3,Lascala4} based on the two-stage
Kalman filter.  Modifications to the transition cost function of this
data association technique to allow for tracking of a
manoeuvring target were also briefly described in \cite{Quach} but
have been found to be difficult to tune, at least in OTHR scenarios.

The structure of the paper is now described.  Section \ref{models}
introduces several classes of dynamical model for manoeuvring targets
which have been used in radar and sonar.  Section \ref{detec} deals
with the problem of manoeuvre detection, that is, estimation of the
onset time of a target manoeuvre based on noisy observations of the
target state. In section \ref{filters} the main methods of manoeuvring
target tracking are summarised. We cover the extension to the case of
cluttered measurements in section \ref{clutter}. In section \ref{conc}
we summarise our observations concerning the manoeuvring target
tracking problem based on the surveyed literature.

\section{Dynamical Models for Manoeuvring Targets}\label{models}
In the following sections we review the literature on manoeuvring
target tracking, focussing first on the case of zero false alarms and
later on the cluttered measurement case.  The starting point for our
survey is a description of dynamical models for manoeuvring targets.
These fall into several broad categories depending on how the
manoeuvre dynamics $x(k)$ are modelled \cite{Barshalom2,Barshalom7}.
We use $y(k)$ to denote the target measurement at time $k$ and $Y^k$
to denote the measurement set to time $k$.  We assume familiarity with
the discrete-time Kalman filter and smoother, and with Bayesian
estimation theory (refer to, for example, \cite{Anderson,Barshalom7}).
\begin{enumerate}
\item 
Systems of the form
\begin{eqnarray}
x(k+1)&=&F(k) x(k)+G(k) u(k) + v(k)\label{M1}\\
y(k)&=&H(k) x(k) + w(k)\nonumber
\end{eqnarray}
where the unknown input (acceleration) process $u(k)$ is a random
process and $v(k)$ is the (small) filter process noise term, which
controls the filter bandwidth.  In this kind of model the target
acceleration $u(k)$ is modelled as an additional process noise term.
Usually a finite-state process is assumed to describe the possible
types of target manoeuvre as is the case in the Markov and semi-Markov
process noise models in which the acceleration switches between several
possible levels according to given transition probabilities and with
given distributions governing the switching times for each level.
\item 
Systems with unknown inputs:
\begin{eqnarray}
x(k+1)&=&F(k) x(k)+G(k) u(k) + v(k)\label{M2}\\
y(k)&=&H(k) x(k) + w(k)\nonumber
\end{eqnarray}
where the  input $u(k)$ is deterministic but unknown
and must be estimated.
Often $u(k)$ is assumed to be piecewise constant, such as in the
multi-level noise case
\begin{eqnarray}
x(k+1)&=&\left\{
\begin{array}{cc}
F(k) x(k)+G(k)u_1(k)+v(k)&\mbox{for model 1}\\
F(k) x(k)+G(k)u_2(k)+v(k)&\mbox{for model 2}\\
\vdots&\vdots\\
F(k) x(k)+G(k)u_r(k)+v(k)&\mbox{for model r}
\end{array}
\right.
\nonumber\\
y(k)&=&H(k) x(k) + w(k)\label{M3}
\end{eqnarray}
where it is assumed that only one of the $r$ possible
target models in effect at any one time. 
\item
Multiple dynamic models
\begin{eqnarray}
x(k+1)&=&F(k;M(k)) x(k)+ v(k;M(k))\label{M4}\\
y(k)&=&H(k;M(k)) x(k) + w(k;M(k))\nonumber
\end{eqnarray}
in which the model type M(k), also called the {\em mode}, is a
finite-state Markov chain taking values in the set
$\{M_1,\ldots,M_r\}$. The transition probabilities of the Markov
chain are usually assumed to be known and the modes are mutually
exclusive.
\end{enumerate}
Systems (\ref{M3}) and (\ref{M4}) are called {\em hybrid systems}
since they contain both continuous-valued states and discrete
parameters (the model type or input level). In the control literature,
system (\ref{M3}) is referred to as a switching control system, and
system (\ref{M4}) a jump-linear system.  In tracking, unlike control
theory, the input term $u(k)$, whether deterministic or random, is
unobservable and must be estimated or allowed for by switching to an
appropriate model.

The particular model structures can be chosen to reflect the expected
target dynamics. This might include models such as constant velocity,
constant acceleration, constant turn-rate and constant speed
(co-ordinated turn), \etc~ The filter process noise $v(k)$ is used to
represent small variations in acceleration about the nominal target
trajectory, while the manoeuvre input $u(k)$, or mode $M(k)$ in the
multiple model case, is used to represent significant levels of
acceleration.  The interested reader is referred to
\cite{Bauschlicher,Watson,Watson1,Alouani1,Blair2} for further examples
of target manoeuvre models.

It is well known that optimal (Bayesian or MAP) state estimation for
hybrid systems generally requires filters with increasing memory
requirements. This has led to several sub-optimal filtering
techniques with bounded complexity such as the interactive
multiple-model algorithm \cite{Blom} and the class of generalised
pseudo-Bayesian algorithms \cite{Ackerson,Chang4}. Readable accounts
of estimation techniques for switching systems include
\cite{Pattipati1}, and \cite{Tugnait}, which covers partitioning
filters, the GPB algorithm and the random sampling algorithm. An IMM
algorithm for semi-Markov switching systems was developed in
\cite{Campo}. We return to a discussion of filters for hybrid systems
in section
\ref{filters}.

\subsection{Correlated Process Noise (Singer) Model}\label{singer}
The archetypal work on modelling a manoeuvring target was done in
\cite{Singer}.  Target manoeuvres are characterised by large
deviations from a constant-velocity trajectory corresponding to a
random acceleration $a(t)$. Thus, in one dimension of motion, if the
components of the state vector $x(t)$ are $x_1(t)$ - the target
displacement and $x_2(t)$ - the velocity, the target dynamics are
described by
\[
\dot{x}(t)=\left[
\begin{array}{cc}
0&1\\0&0
\end{array}
\right]
x(t)+\left[
\begin{array}{c}
0\\1
\end{array}
\right]
a(t).
\]
Manoeuvres are parametrised by an acceleration variance
$\sigma_m^2$ and a time constant $1/\alpha$.
During a manoeuvre, the target acceleration is {\em correlated}
with autocorrelation function
\[
r(\tau)={\rm E}\{a(t)a(t+\tau)\}=\sigma_m^2 {\rm e}^{-\alpha|\tau|}.
\]
The same autocorrelation results from a linear system
\[
\dot{a}(t)=-\alpha a(t)+v(t)
\]
driven by white noise $v(t)$ with autocorrelation function $2\alpha
\sigma_m^2 \delta(t)$ ($\delta(\cdot)$ is the Kronecker delta function).
Augmenting the state with the target acceleration $a(t)$,
the dynamical model for the manoeuvring target becomes
\begin{equation}\label{sing}
\dot{x}(t)=\left[
\begin{array}{ccc}
0&1&0\\
0&0&1\\
0&0&-\alpha
\end{array}
\right]
x(t)+\left[
\begin{array}{c}
0\\0\\1
\end{array}
\right]
v(t)
\end{equation}
where $v(t)$ is white noise with variance $2\alpha\sigma_m^2$.

The continuous-time model can be discretised in the usual manner 
\cite{Astrom} by uniform sampling with period $T$. The resulting
discrete-time model is (with a slight abuse of notation)
\begin{eqnarray}
x(k)&=&\left[
\begin{array}{l}
\mbox{target displacement at time $kT$}\\
\mbox{target velocity at time $kT$}\\
\mbox{target acceleration at time $kT$}
\end{array}
\right]\nonumber\\
x(k+1)&=&F_{\alpha}(T) x(k)+u(k)\label{man1}
\end{eqnarray}
where the transition matrix $F_{\alpha}(T)$ may be shown to be
\begin{equation}\label{SingerF}
F_{\alpha}(T)=\left[
\begin{array}{ccc}
1&T&(\alpha T-1+{\rm e}^{-\alpha T})/\alpha^2\\
0&1&(1-{\rm e}^{-\alpha T})/\alpha\\
0&0&{\rm e}^{-\alpha T}
\end{array}
\right]
\end{equation}
and where $u(k)$ is a zero-mean white noise sequence
with covariance matrix $Q(k)$ whose elements $q_{ij}$ are given by
\begin{eqnarray}
q_{11}&=&\sigma_m^2[1-{\rm e}^{-2\alpha T}+
2\alpha T+\frac{2}{3}\alpha^3T^3-2\alpha^2T^2-4\alpha T{\rm e}^{-\alpha T}
]/\alpha^4\nonumber\\
q_{12}&=&\sigma_m^2[{\rm e}^{-2\alpha T}
+1-2{\rm e}^{-\alpha T}+2\alpha T {\rm e}^{-\alpha T}-2\alpha T+\alpha^2T^2
]/\alpha^3\nonumber\\
q_{13}&=&\sigma_m^2[1-{\rm e}^{-2\alpha T}
-2\alpha T {\rm e}^{-\alpha T}]/\alpha^2\nonumber\\
q_{22}&=&\sigma_m^2[4{\rm e}^{-\alpha T}
-3-{\rm e}^{-2\alpha T}+2\alpha T]/\alpha^2\nonumber\\
q_{23}&=&\sigma_m^2[{\rm e}^{-2\alpha T}+1-2{\rm e}^{-\alpha T}]/\alpha\nonumber\\
q_{33}&=&\sigma_m^2[1-{\rm e}^{-2\alpha T}]/\alpha. \label{SingerQ}
\end{eqnarray}
The manoeuvre parameter $\alpha$ may be chosen to model various
classes of target manoeuvre, \eg, lazy turn, evasive manoeuvre, \etc~ 
If the sampling time is sufficiently small compared with the manoeuvre
parameter, \ie, $\alpha T \ll \half$, the following asymptotic form of
$Q(k)$ may be used:
\begin{equation}
\bar{Q}(k)=2\alpha\sigma_m^2\left[
\begin{array}{ccc}
T^5/20&T^4/8&T^3/6\\
T^4/8&T^3/3&T^2/2\\
T^3/6&T^2/2&T
\end{array}
\right].
\end{equation}
If noisy measurements of the target are available, {\em viz.}
\[
y(k)=H(k) x(k)+w(k),~w(k)\sim N\{0,R(k)\},
\]
where $N\{m,C\}$ denotes a Gaussian density\footnote{We will also use
$N\{y;m,C\}$ to denote the value at $y$ of the Gaussian density with
mean $m$ and covariance $C$.} with mean $m$ and covariance $C$, then
it is straightforward to derive a Kalman filter for minimum
mean-square error estimation of the manoeuvring target's state (see
section \ref{filters}).

The Singer model has been extended in \cite{Helferty} to reflect the
tendency of targets to have a constant forward speed with acceleration
normal to their velocity. Under this model the turn rate is assumed to
be uniformly distributed. A higher order ``jerk model'' of
manoeuvring targets has been developed in \cite{Mehrotra}. This model
is obtained by discretising a system similar to (\ref{sing}) the state
of which contains a third-order derivative of the position. It is
claimed that this model more accurately reflects the behaviour of
agile targets.  The filter for this fourth-order model requires
three-point initiation with the initial jerk estimate taken as zero.

\subsection{Semi-Markov Correlated Process Noise Model}
\begin{figure}[h]
\centerline{\psfig{figure=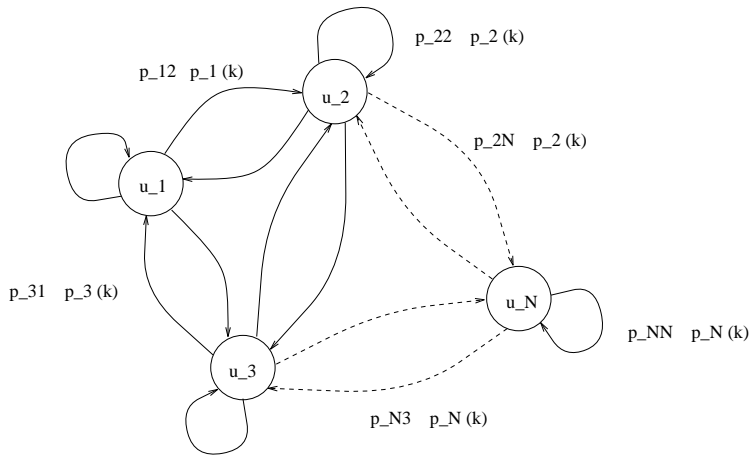,height=6cm}}
\caption{\protect{\small Finite-state, discrete-time semi-Markov model
of manoeuvring target acceleration. The model parameters are the
transition probabilities $p_{ij}$ and the holding-time distributions
$p_i(k)$ which determine the probability of switching as a function
of the time spent in the current manoeuvre state.}}
\label{Fig2}
\end{figure}

The correlated process noise model of Singer was built on in
\cite{Moose2,Moose5,Gholson,Ricker}.
By allowing the correlated process noise to have a randomly switching
mean value, rather than a zero mean value, a more realistic model of a
manoeuvring targets was obtained. This model is also more economical
than the white Gaussian process noise with randomly switching mean
since it requires fewer mean values for accurate modelling than the
latter.
The system model for a target with state $x(k)$ and
measurement process $y(k)$ is in this case
\begin{eqnarray}
x(k+1)&=&F(k) x(k)+G(k)(u(k)+v(k))\label{man2}\\
y(k)&=&H(k) x(k) + w(k)\nonumber
\end{eqnarray}
where $u(k)$ is a random process taking values in a finite, discrete
set $\{u_1,\ldots,u_N\}$; $v(k)$ is the correlated Gaussian (Singer)
process noise; and $w(k)$ is a white Gaussian measurement noise. The
Singer process noise covariance and the measurement noise covariance
are assumed to be known. Since physical target manoeuvres tend to be
sustained for a certain ``holding'' time, the randomly switching mean
is modelled as a {\em semi-Markov} process \cite{Howard} (Vol. 2), and
this is illustrated in Fig. \ref{Fig2}.  In addition to the transition
probabilities for a (time-invariant) Markov chain
\[
p_{ij}=\Pr\{u(k+1)=u_j|u(k)=u_i\},
\]
a further set of probabilities, the holding-time distributions, must also
be specified:
\[
p_i(k)=\Pr(\mbox{transition from $u_i$}\,|\,\mbox{$k$ successive time instants 
spent in $u_i$}).
\]

To simplify the characterisation of the semi-Markov process, the
holding-time distribution is often assumed to be an monotonically
increasing (\eg, exponential) function of time so that a transition
from manoeuvre state $u_i$ becomes more likely as the time spent in
$u_i$ increases. In practice, to reduce the number of free parameters
in the model, it may be assumed that each possible manoeuvre state is
held for a random holding-time, with the transition matrix of the
Markov chain having diagonal elements near unity and equal
off-diagonal elements \cite{Moose2}.

Minimum variance (conditional mean) state estimation of the
manoeuvring target system (\ref{man2}) requires a filter with
exponentially increasing memory \cite{Ackerson}. Approximate conditional
mean filtering is therefore necessary and this take the form of a
truncated, weighted sum of Kalman filter estimates matched to the
possible manoeuvre sequences. By forcing the covariance update for each
filter in the bank to be identical, it is possible to reduce the
filter complexity to that of a Kalman filter with $N$ state updates
\cite{Farina,Moose5} (see section \ref{opt1}).

\subsection{Poisson Processes and Renewal Models}
Target manoeuvres have been modelled as a finite-state jump process
with Markov transitions \cite{Lim}. The theory for these
models is that of discrete-state, continuous-time Markov processes
\cite{Bharucha} and is beyond the scope of this article.
We will however summarise the main points of these alternative
approaches to manoeuvre modelling and the different filter structures
which arise.

\begin{figure}[h]
\centerline{\psfig{figure=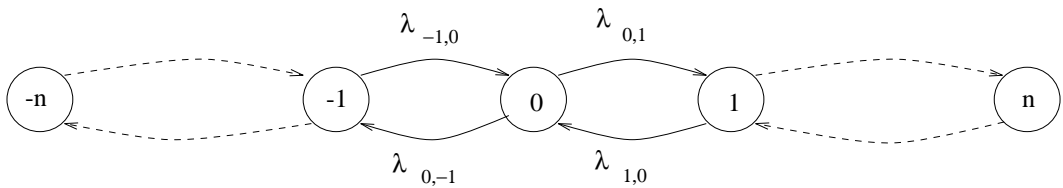,width=14cm}}
\caption{\protect{\small Discrete-state Poisson process manoeuvre
model.}}
\label{Fig3}
\end{figure}

Consider the case of motion in one dimension represented by the
discrete-time system
\begin{eqnarray*}
x(k+1)&=&F x(k)+ G(u(k)+v(k)),~v(k)\sim N\{0,Q\}\\
y(k)&=&H x(k) + w(k),~w(k)\sim N\{0,R\}
\end{eqnarray*}
where $u(k)$ is a scalar process corresponding to the jumps from one
manoeuvre state to another, and the other symbols have their usual
meanings.  Let the possible manoeuvre states, shown in
Fig. \ref{Fig3},  be denoted $\{0,\pm 1,
\pm 2,\ldots, \pm n\}$, with state 0 corresponding to a
constant-velocity target. Note that transitions are only allowed
between adjacent manoeuvre states, and the transition rates
$\lambda_{ij}$ (probability per unit time) between states $i$ and $j$
are assumed to be known constants.  If $p_i(k)=\Pr\{u(k)=i\}$ denotes
the probability that the target is in manoeuvre state $i$ at time $k$,
then for $i\neq \pm n$,
\[
p_i(k+\Delta)=\lambda_{i-1,i}\Delta p_{i-1}(k)+(1-\lambda_{i,i+1}\Delta
-\lambda_{i+1,i}\Delta) p_{i}(k)+\lambda_{i+1,i}\Delta p_{i+1}(k)
\]
where $\Delta$ is a small time increment. Similar recursions hold for
the two states at the end of the chain $i=\pm n$.  A (constrained)
linear, minimum variance filter for the state estimates is then
derived based on noisy observations. The state and covariance update
for the filter are the same as those in the Kalman filter, but the
prediction equations for the state and covariance are different. For
example, the state prediction is
\[
\xhat(k+1|k)=F \xhat(k|k)+G b(k+1)
\]
where the term $b(k+1)$ is the predicted acceleration from the Poisson
process, and may be computed using the transition rates $\lambda_{ij}$
and posterior manoeuvre state probabilities $\Pr(u(k)=i|Y^k)$.  The
probabilities $\Pr(u(k)=i|Y^k)$ presumably satisfy a recursion,
although this is not provided in \cite{Lim}. The covariance prediction
is of the form
\[
P(k+1|k)=F P(k|k) F' + G Q G'+ G B G'
\]
where $B$ is a diagonal matrix determined from the $\Pr\{u(k)=i\}$
and the $\Pr(u(k)=i|Y^k)$ probabilities.

No comparison with the classical manoeuvre modelling techniques is
provided. More importantly, it is unclear how robust the filter is
when the transition rates of the manoeuvre process are not precisely
known.

Finite-state renewal process models of target manoeuvres have been
investigated in \cite{Sworder2}. Renewal processes \cite{Cox} are
generated by the differences between consecutive transition times (or
sojourn times) of a point process. For example, if
$\{t_1,\ldots,t_n,\ldots\}$ is a Poisson process, then
$\{\tau_1=t_1,\tau _2=t_2-t_1,\ldots,\tau_n=t_n-t_{n-1},\ldots\}$ is a
renewal process \cite{Papoulis}.  Renewal processes, which are Markov
processes, are characterised up to an initial distribution by a
matrix of transition probabilities ${\bf P}$ between the manoeuvre
states $\{u_1,\ldots,u_N\}$, and a set of sojourn-time distributions
$p_i(\tau|u(k)=u_i)$. Thus both the transitions between manoeuvre
states and the time spent in each state following a transition are
random.

A general characteristic of Markov and semi-Markov manoeuvring target
models is that the sojourn times (\ie, the time spent in each
manoeuvre state) are exponentially distributed.  Thus the sample paths
from such processes can contain manoeuvring segments of very short
duration, which may not be physically realistic.

A model for an agile target that executes co-ordinated (constant turn
rate at constant speed) turns at instants governed by a renewal
process in developed in \cite{Sworder2}.  A dynamical model for
constant-turn rate motion in the $(x_1,x_2)$ plane is
\cite{Barshalom7}
\[
\ddot{x_1}=-\omega \dot{x_2},~\ddot{x_2}=-\omega \dot{x_1}
\]
where $\omega$ is the (constant) angular velocity. The state-space model
for the manoeuvring target is then\footnote{Strictly speaking, we should
write this as a stochastic differential equation since the derivative
of the Brownian motion term $v$ is not defined.}
\begin{equation}\label{turn}
\left[
\begin{array}{c}
\dot{x_1}(t)\\
\dot{x_2}(t)\\
\ddot{x_1}(t)\\
\ddot{x_2}(t)
\end{array}
\right]=\left[
\begin{array}{cccc}
0&0&1&0\\
0&0&0&1\\
0&0&0&-\omega(t)\\
0&0&\omega(t)&0
\end{array}
\right]\left[
\begin{array}{c}
x_1(t)\\
x_2(t)\\
\dot{x_1}(t)\\
\dot{x_2}(t)\\
\end{array}
\right]+\left[
\begin{array}{cc}
0&0\\0&0\\1&0\\0&1
\end{array}
\right]v(t)
\end{equation}
where $v(t)$ is a white noise process and $\omega(t)$ a finite-state renewal
process modelling the random rate and duration of the turns.

The transition probabilities of the renewal process are assumed known.
Each sojourn-time distribution is characterised by Gamma density with
known parameters
\[
p_{\gamma}(t)=\left\{
\begin{array}{ll}
\frac{\lambda (\lambda t)^{r-1}}{\Gamma(r)}\exp(-\lambda t)&t>0\\
0&\mbox{otherwise}
\end{array}
\right.
\]
where $r>0$ controls the shape of the density and $\lambda>0$ governs
the time scale. Notice that for $r=1$, the Gamma density reduces to an
exponential PDF, but for $r>0$ the Gamma density function is zero at
the origin, so that very short sojourns in the manoeuvre states are uncommon.

The model (\ref{turn}) is not well suited to extended Kalman filtering
since it is has non-Gaussian noise multiplying the state components.
The dynamical process can be approximated by a Gaussian process with
power spectral density (PSD) matched to that of the equilibrium PSD of
the non-stationary process generated by (\ref{turn}).  Analogously to
the Singer case, a state-space system augmented by the manoeuvre
$\omega(t)$ process is obtained. A second order matching to the PSD is
required for a reasonable approximation to the original system, and
depends on the time constant of the matching filter and the turn-rate
variance ${\rm E}\{\omega^2\}$. The dimension of the equivalent
continuous-time Gauss-Markov system is then six and this system is
amenable to extended Kalman filtering.  Results of simulations
performed in \cite{Sworder2} indicate that the performance improvement
from this matched EKF method is not commensurate with its increased
complexity for single-sensor systems. However, the performance gain is
significant for image-enhanced systems \cite{Sworder1} such as a
microwave radar teamed with an infra-red imaging sensor. It should be
stressed that the renewal model of manoeuvres was developed for highly
agile targets which are not well modelled by the conventional semi-Markov
approach.

\section{Manoeuvre Detection}\label{detec}
Many manoeuvring target tracking techniques require a method for
detecting the manoeuvre before compensation of the state estimate can
be accomplished. This is the case in approaches which use a
``quiescent'' constant-velocity model such as the variable-dimension
filter \cite{Barshalom2,Barshalom11} which uses a low order Kalman
filter to provide high accuracy tracking of nearly constant velocity
targets, and switches to a higher order filter once the manoeuvre is
detected. Manoeuvre detection is also used in an adaptive
Kalman filter to switch to a higher level of process noise during a
manoeuvre.  The input estimation technique
\cite{Chan} also requires knowledge of the manoeuvre onset time.
The IMM algorithm (section \ref{opt2}) and its variants do not require
separate manoeuvre detection logic.

Manoeuvre detection can be formulated as a hypothesis testing problem
and implemented by likelihood ratio test (LRT) based on the Kalman
filter innovations. The central idea is that the sequence of Kalman
filter innovations are white Gaussian random variables for a correctly
modelled system. The probability of the measurement sequence is then
expressible solely in terms of the innovations and their covariances.
The statistics of the Kalman filter innovations can therefore be
checked to see if they are consistent with the hypothesised model of
the target dynamics.

Likelihood ratio tests and other forms of hypothesis testing are
described in detail in \cite{Vantrees1}.  The likelihood ratio test
(LRT) arises from the Neyman-Pearson criterion and uses (i) a test
statistic, and (ii) a threshold for the test.  If the test statistic,
which is a function of the measurement data, exceeds the threshold, a
manoeuvre is declared; otherwise no manoeuvre is declared. Assuming
that the probability distribution (of the test statistic) under the
manoeuvre and non-manoeuvre hypotheses are known, fixing a value for
the probability of false alarm $P_{MFA}$ determines the threshold for
the test.  The probability of manoeuvre detection $P_{MD}$
\cite{Thorp} and can then be calculated for the chosen
threshold. There is clearly a compromise since we cannot make $P_{MD}$
arbitrarily large without also raising $P_{MFA}$.

The following description of an innovations-based manoeuvre detector
is from \cite{Barshalom2}.  If $\nu(k)$ denotes the Kalman filter innovation,
\ie, the difference between the measurement and its prediction, and
$S(k)$ is the corresponding innovations covariance matrix, then the
normalised squared innovation is defined as
\begin{equation}
\epsilon(k)=\nu'(k) S^{-1}(k) \nu(k).
\end{equation}
In a correctly tuned KF with Gaussian inputs,  $\epsilon(k)$ is
a $\chi^2_{n_y}$ random variable where $n_y$ is the number of
measurement components. The modified likelihood function of the Kalman
filter is given by
\begin{equation}
\lambda(k)=\sum_{i=1}^k\epsilon(k)
\end{equation}
and forms the test statistic for manoeuvre detection.
This function can be made responsive to recent data by applying
an ``exponential forgetting factor'' \cite{Ljung} yielding
the fading-memory likelihood function
\begin{equation}
\rho(k)=\sum_{i=1}^k \alpha^{k-i}\epsilon(i)
\end{equation}
which satisfies the recursion
\begin{equation}
\rho(k)=\alpha \rho(k-1)+\epsilon(k).
\end{equation}
The forgetting factor $\alpha$ (not the same as in the Singer model)
is between 0 and 1, and yields an effective memory length of
$1/(1-\alpha)$.  Equivalently, the modified likelihood function can be
implemented as a sliding window sum. The mean value of $\rho(k)$ can be
shown to be $n_y/(1-\alpha)$ \cite{Barshalom2}. The detection test
compares $\rho(k)$ to a threshold to determine whether or not a manoeuvre has
occurred. Similarly, a hypothesis test on the significance of
the acceleration estimate from the manoeuvring Kalman filter is used
to determine when the manoeuvre has terminated.

It is important to detect manoeuvres as soon as they occur,
but there is a trade-off between the detection delay and the
probability of falsely declaring that a manoeuvre has occurred.
In \cite{WangTC} a method was proposed to minimise the
delay in manoeuvre detection. Their technique uses a sliding window
over which the Kalman filter residuals are summed:
\[
D_L(k)=\sum_{i=k-L+1}^k \nu(i).
\]
The random variable $D_L(k)$ is normally distributed with zero mean
for a constant-velocity target but has a non-zero mean if the target
is accelerating (the mean can be calculated from the system matrices
and the target acceleration over the window).  For a given false alarm
probability $P_{FA}$ it is possible to compute the average manoeuvre
detection delay, which depends on the system matrices, the target
acceleration, $P_{FA}$ and window length $L$. With all parameters
except $L$ fixed, it is then possible to determine numerically the
optimum window length that minimises the manoeuvre detection delay.

In tracking and other practical scenarios, the probability densities
for hypothesis testing may not be precisely known, rendering it
difficult to determine a test statistic and threshold for the
likelihood ratio test. Moreover, incorrect modelling of the required
probability densities decreases the power of the test, \ie, its
ability to detect a manoeuvre when one has occurred.  

The same remarks apply to generalised likelihood ratio tests (GLRT's)
where the probability densities in question depend on an unknown
parameter (\eg, the acceleration) or composite hypothesis. In a GLRT,
maximum likelihood estimates of the parameter are used in a LRT to
make a decision on which hypothesis is true.
An example of an application of a GLRT in manoeuvre detection
may be found in \cite{McAulay,Korn} and is summarised in \cite{Farina}.

If it is desired to estimate both the onset time of the manoeuvre and the
target acceleration at this time, the following hypothesis test on the
Kalman filter innovation sequence can be set up
\begin{eqnarray*}
H_1:&&\nu(k)=\nu_0(k)+Gu(\tau)\\
H_0:&&\nu(k)=\nu_0(k)
\end{eqnarray*}
where the acceleration $u(\tau)$ is zero for $k<\tau$ and constant
thereafter, and $\nu_0(k)$ is the residual for a constant-velocity target.
A sufficient statistic for the test is
\[
L=\frac{{\rm p}(\nu(k)|Y^k,H_1)}{{\rm p}(\nu(k)|Y^k,H_0)}
\]
which is implemented by discretising the possible accelerations
and onset times $u$ and $\tau$ and passing the filter innovation
through a bank of low-pass filters matched to the possible combinations
of $u$ and $\tau$. The optimum estimate of $u$ and $\tau$ corresponds
to the filter with the largest magnitude output. Although this technique
estimates both the acceleration and onset time of the manoeuvre, the detection
is delayed by the memory of the low-pass filter, and the complexity of the
algorithm is substantial.

Several techniques exist in the literature for designing more robust
detection tests, these include the marginalised likelihood ratio test
(MLRT) \cite{Gustafson} and a robust LRT reported in \cite{White}. It
seems that these techniques have so far not been applied in
manoeuvring target tracking. The MLRT is designed for systems
undergoing sudden changes when the onset time and magnitude of the
change are unknown. It uses marginalisation to remove the dependence
of the probability densities on the parameters and also does not
require a threshold. Despite this advantage, the MLRT has a high
complexity and also requires data smoothing.  The robust LRT in
\cite{White} focusses mainly on developing a hypothesis test for
nonlinear dynamical systems in low signal-to-noise ratio conditions.

\section{Manoeuvring Target Tracking Techniques}\label{filters}
In this section we review the existing techniques for tracking a
manoeuvring target in the absence of clutter.  The literature on this
subject is replete with claims of superior performance; however, many
simulation studies are unconvincing or fail to take account of
existing approaches in their performance comparisons.  The general
absence of accepted tests for manoeuvring and multi-target tracking
has been noted in \cite{Blair5}.  Recently a benchmark problem for
manoeuvring target tracking with a phased-array microwave radar was
established \cite{Blair5}. This benchmark problem requires the tracker
to initiate and maintain track on manoeuvring targets within a minimum
number of scans (on average), subject to a maximum track loss
probability of 4\%. Test targets are subject to position and
manoeuvrability constraints while fading, missed detections, beam
shape and resolution are taken into account.

The Kalman filter is the optimal (MMSE) state estimator for linear
observations of a target with known dynamics and white Gaussian
process and measurement noise. However, as soon as a target changes
its dynamics, \eg, by executing a manoeuvre, the Kalman filter
estimates may diverge causing track loss. This divergence is due to
incorrect modelling of the target dynamics, rather than finite
numerical precision, and is the chief difficulty in manoeuvring target
tracking. As summarised in \cite{Farina}, the approaches to
manoeuvring target tracking using Kalman filters generally fall into 5
classes:
\begin{enumerate}
\item 
re-initialisation of the Kalman gain (to reduce the
effect of past measurements);
\item 
increasing the process noise covariance;
\item 
increasing the state estimation covariance;
\item 
augmenting the filter state with acceleration terms
(variable-dimension filter);
\item 
switching to or applying heavier weighting to another
filter in a bank of Kalman filters (multiple-model approaches).
\end{enumerate}
Once a manoeuvre is detected, it becomes necessary to take one of the
actions above in order to correct the state estimate and its
covariance.  Approaches 1--3 above all involve changing the {\em
parameters} of the Kalman filter.  For a given application, the
adjustment required will depend on the type of manoeuvre, the sampling
time, the measurement noise, \etc~ Usually, trial-and-error methods
must be used to tune the Kalman filter for manoeuvre tracking.
Approaches 4 and 5 involve changing or adapting the {\em filter
structure}, and we describe these later in some detail. The following
sections cover the major filtering techniques and structures for
manoeuvring target tracking, including: optimal Bayesian filters for
multi-level (Markov) process noise and multiple models; single-scan
and multiple-scan approximate filters including the interacting
multiple-model (IMM) algorithm and the generalised pseudo-Bayesian
(GPB) filter; input estimation; the variable dimension filter; the
two-stage Kalman filter (which treats the acceleration as a bias
term); the IMM smoother; and an expectation-maximisation approach.
The multi-hypothesis tracker (MHT) \cite{Reid} is also of importance
in manoeuvring target tracking, and we comment on this in 
section \ref{mht}.

\subsection{Optimal and Approximate Filters for Markov Noise}\label{opt1}
\begin{figure}[h]
\centerline{\psfig{figure=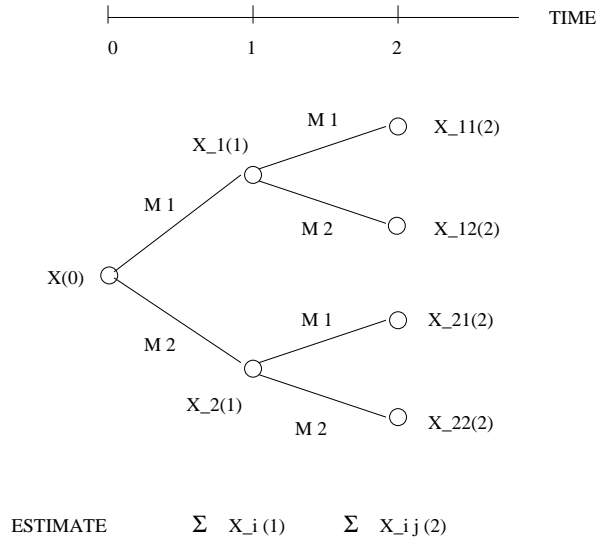,height=7cm}}
\caption{\protect{\small Idea behind branching filters (\eg, MHT) and optimal Bayesian
algorithms with two possible manoeuvre models $M1$ or $M2$ at each
time instant.  The MAP estimate is obtained by taking the most
probable manoeuvre sequence at time $k$, while the MMSE
estimate corresponds to a weighted sum of all conditional state
estimates at a given time. Sub-optimal approximations 
are obtained by merging common path histories and deleting unlikely
branches.}}
\label{Fig4}
\end{figure}

The optimal (Bayesian) filter can be derived for tracking a
manoeuvring target whose acceleration $u(k)$ is represented by a
finite-state Markov chain with state space $\{u_1,\ldots,u_N\}$ and
known transition probabilities $p_{ij}$. The system model is, as in
equation (\ref{M3}),
\begin{eqnarray*}
x(k+1)&=&F x(k) + Gu(k) + G v(k),~v(k)\sim N\{0,Q(k)\}\\
y(k)&=&H x(k) + w(k),~w(k)\sim N\{0,R(k)\}
\end{eqnarray*}
where the state $x(k)$ contains position, velocity and acceleration
terms with corresponding $F$ matrix (\ref{SingerF}) and $v(k)$ is the
zero-mean Singer noise component with covariance $Q$ given by
(\ref{SingerQ}). In this derivation we will not use the semi-Markov nature
of the manoeuvre process $u(k)$. As in \cite{Farina}, we denote the $j$th
sequence of possible target manoeuvres by 
\[
\Omega_j^k=\{u_{j_1}, \ldots, u_{j_k}\}.
\]
From Bayes' rule, the conditional mean estimate of $x(k)$ based on
the measurements contained in $Y^k$ is expressible as
\begin{equation}\label{optimal}
\xhat(k|k)=\sum_{i=1}^{N^k} \, \Pr(\Omega_i^k|Y^k)\xhat_i(k|k)
\end{equation}
with conditional estimates defined as $\xhat_i(k|k)={\rm
E}\{x(k)|\Omega_i^k,Y^k\}$. Alternatively, we could take the MAP
estimate, \ie, the $\xhat_i(k|k)$ for which $\Pr(\Omega_i^k|Y^k)$ is
maximum. In either case, the optimal estimate is seen to require the
evaluation of $N^k$ terms which implies that the optimal tracker has
exponentially increasing computational requirements, as
illustrated in Fig. \ref{Fig4}.

A natural way of reducing the complexity of the filter is to approximate the
$N^{k-1}$-term mixture PDF of the state at time $k-1$ by a single Gaussian
density, thus obtaining a filter with $O(N)$ complexity \cite{Moose2}:
\begin{eqnarray*}
\xhat(k|k)&\approx&\sum_{i=1}^N c_i(k) \,\xhat_i(k|k)\\
c_i(k)&=&\Pr(u(k)=u_i|Y^k)
\end{eqnarray*}
where $\xhat_i(k|k)={\rm E}\{x(k)|u(k)=u_i,Y^k\}$, which is generally
not the same as ${\rm E}\{x(k)|\Omega_i^k,Y^k\}$. The conditional
state estimates are obtained by Kalman filtering in the usual manner:
\begin{eqnarray*}
\xhat_i(k|k)&=&\xhat_i(k|k-1)+W(k)[y(k)-\yhat_i(k|k-1)\\
\xhat_i(k|k-1)&=&F\xhat_i(k-1|k-1)+G u_i\\
\yhat_i(k|k-1)&=&H\xhat_i(k|k-1)\\
W(k)&=&P(k|k-1)H'S^{-1}(k)\\
S(k)&=&H P(k|k-1) H'+R(k)\\
P(k|k-1)&=&F P(k-1|k-1) F'+G Q(k) G'\\
P_i(k|k)&=&(I-W(k)H)P(k|k-1)
\end{eqnarray*}
where $P_i(k|k)$ is the error covariance associated with $\xhat_i(k|k)$.
The net covariance of the state estimate is obtained using a standard
formula for Gaussian mixtures \cite{Barshalom7}
\begin{eqnarray*}
P(k|k)&=&\sum_{i=1}^N c_i(k)\{P_i(k|k)+\xhat_i(k|k)\xhat_i(k|k)'\}-\xhat(k|k)\xhat(k|k)'\\
&=&(I-W(k)H)P(k|k-1)-\sum_{i\neq j} c_i(k) c_j(k) \xhat_i(k|k)\xhat_j(k|k)'
\end{eqnarray*}
where we have used the fact that the $c_i(k)$ sum to unity, and the assumption
that the noise covariances $Q$ and $R$ are identical across all manoeuvre
modes. To evaluate the
posterior manoeuvre probabilities $c_i(k)$, a recursion can be derived
using Bayes' rule:
\begin{eqnarray*}
c_i(k)&=&\Pr(u(k)=u_i|y(k),Y^{k-1})\\
&=&\Delta^{-1} {\rm p}(y(k)|u(k)=u_i,Y^{k-1})\Pr(u(k)=u_i|Y^{k-1}),
\end{eqnarray*}
writing the normalising constant as $\Delta^{-1}$. Furthermore the above
terms are by assumption
\[
{\rm p}(y(k)|u(k)=u_i,Y^{k-1})=N\{\yhat_i(k|k-1),S(k)\}
\]
and
\[
\Pr(u(k)=u_i|Y^{k-1})=\sum_{j=1}^N
\Pr(u(k)=u_i|u(k-1)=u_j,Y^{k-1})\Pr(u(k-1)=u_j|Y^{k-1})
\]
so that the desired recursion is
\[
c_i(k)=\Delta^{-1} N\{\yhat_i(k|k-1),S(k)) \sum_{j=1}^N p_{ji} c_j(k-1).
\]
Finally the state estimate update can be written
\begin{eqnarray*}
\xhat(k|k)&=&\sum_{i=1}^N c_i(k) \xhat_i(k|k)\\
&=&\sum_{i=1}^Nc_i(k)\xhat_i(k|k-1)+\sum_{i=1}^Nc_i(k)Gu_i\\
&=&(I-W(k)H)(F\xhat(k-1|k-1)+G \uhat(k))+W(k)y(k)
\end{eqnarray*}
where $\uhat(k)=\sum_{i=1}^Nc_i(k) u_i$ is the estimated target
acceleration at time $k$.
Notice that the Kalman gain and covariances for the conditional
state estimate are identical, which greatly simplifies the filter
update.

\subsection{Input Estimation}
The input estimation technique treats the target acceleration as
deterministic bu unknown vector. The ensuing description of the
technique is from \cite{Barshalom2} and is based on \cite{Chan}. Assume
a system model of the form
\begin{eqnarray}
x(k+1)&=&F x(k) + Gu(k) + v(k),~v(k)\sim N\{0,Q(k)\}\label{ipest}\\
y(k)&=&H x(k) + w(k),~w(k)\sim N\{0,R(k)\}\nonumber
\end{eqnarray}
for known matrices $F$, $G$, $H$ and noise covariances $Q$, $R$. The
acceleration sequence $\{u(l)\}$ is assumed to be zero for $l<k$, and
non-zero for $k\leq l < k+s$.  Denote the state estimates from the
constant-velocity $u(k)\equiv 0$ target model as $\xhat_0(k|k)$
and those for (\ref{ipest}) as $\xhat(k|k)$.
The prediction of these estimates is accomplished via
\begin{equation}\label{prop}
\xhat_0(i+1|i)=F(I-W(i)H)\xhat_0(i|i-1)+F W(i) y(i).
\end{equation}
Denote the estimator transition matrix by $\Phi(i)=F(I-W(i)H)$
and the $i$-step transition matrix by
\[
\Phi(k,i)\keiko\prod_{j=k}^i \Phi(j)=\Phi(i)\Phi(i-1)\cdots\Phi(k)
\]
Starting with $\xhat_0(k|k-1)=\xhat(k|k-1)$, and propagating the difference
equation (\ref{prop}) \cite{Kailath}
we obtain the constant-velocity
predictions for $i=k,\ldots,k+s-1$ as
\begin{equation}
\xhat_0(i+1|i)=\Phi(k,i)\xhat(k|k-1)+\sum_{j=k}^i \Phi(k,j-1) FW(j)y(j),
\end{equation}
whereas the correct predictions are in fact
\[
\xhat(i+1|i)=\Phi(k,i)\xhat(k|k-1)+\sum_{j=k}^i \Phi(k,j-1)(FW(j)y(j)+Gu(j))
\]
with innovations $\nu(i+1)=y(i+1)-H\xhat(i+1|i)$. The innovations are
white, zero-mean and have covariance $S(i+1)$ (the innovations
covariance in the Kalman filter for (\ref{ipest})). The innovations sequence
of the mistuned (zero-velocity) filter is
\[
\nu_0(i+1)=y(i+1)-H\xhat_0(i+1|i).
\]
By assuming that the accelerations $u(j)=u$ are constant over the interval
$[k,k+s-1]$, we have that
\[
\nu_0(i+1)=\Psi(i+1) u + \nu(i+1),~\Psi(i)=H\sum_{j=k}^i\Phi(k,j-1)G
\]
which results in a standard least-squares problem for the unknown
input vector $u$. Now write this as
\[
\nu_0=\Psi u +  \nu,~\nu\sim N\{0,S\}
\]
where $\nu=[\nu(k+1),\ldots,\nu(k+s)]'$, $S={\rm blockdiag}(S(i))$, \etc~
The least-squares estimate for $u$ is then
\[
\uhat=(\Psi'S^{-1}\Psi)^{-1}\Psi'S^{-1}\nu_0
\]
and has mean-square error $P_u=(\Psi'S^{-1}\Psi)^{-1}$. The same
solution can of course be obtained by recursive least-squares. The
acceleration estimate $\uhat$ can be employed to test the manoeuvring
hypothesis, with a manoeuvre declared if
\[
d(\uhat)=\uhat'P_u^{-1}\uhat\geq c
\]
for some threshold $c$ chosen from a $\chi^2_{n_u}$ table for a given
false-alarm probability ($n_u$ is the dimension of the input
vector).

When a manoeuvre is declared, the constant-velocity state estimate
and its covariance are corrected according to
\begin{eqnarray*}
\xhat(k+s+1|k+s)&=&\xhat_0(k+s+1|k+s)+M \uhat\\
P(k+s+1|k+s)&=&P(k+s+1|k+s)+M P_u M'
\end{eqnarray*}
where $M=\sum_{j=k}^i\Phi(k,j-1) G$. The manoeuvre is declared finished
when the correction becomes insignificant. The key parameter in the algorithm
is the window length $s$, the choice of which depends on the sampling interval.

Various modifications of the input estimation technique have appeared
\cite{Farooq1,Bogler,Park}. The technique does not require
retrospective corrections to the estimate, and reportedly yields
reasonable performance \cite{Farooq2}, but at a relatively high
complexity.  Input estimation was compared with the IMM filter in
\cite{Barshalom12,Farooq3,Hou}. The assumption of known manoeuvre
onset time is addressed in \cite{Bogler} where a bank of filters was
used to estimate the onset time of the manoeuvre--- increasing the
complexity of the technique by at least an order of magnitude.  It is
claimed \cite{Farooq1} that enhanced performance can be obtained at
significantly reduced complexity using the information form of the
Kalman filter \cite{Anderson}.  In \cite{Bogler} it is suggested that
the technique is well-suited to implementation in the framework of a
multi-hypothesis tracker (see section \ref{clutter}).

\subsection{Variable-Dimension Filter}\label{vardim}
The idea behind variable-dimension filtering was already mentioned in section
\ref{detec}. A constant-velocity model for the target is adopted
during quiescent (non-manoeuvring) tracking periods. Manoeuvre detection is
accomplished by monitoring a fading-memory average of the normalised
residuals from the Kalman filter. Once a manoeuvre is detected, the
constant-velocity model is abandoned and the algorithm ``switches'' to an
augmented model such as the one-dimensional, mean-jerk acceleration
(Wiener process noise) model:
\[
x(k+1)=F x(k) + G v(k),~v(k)\sim N\{0,Q(k)\}
\]
\begin{equation}\label{wiener}
F=\left[
\begin{array}{ccc}
1&T&T^2/2\\
0&1&T\\
0&0&1
\end{array}
\right],~
Q=\left[
\begin{array}{ccc}
T^5/20&T^4/8&T^3/6\\
T^4/8&T^3/3&T^2/2\\
T^3/6&T^2/2&T
\end{array}
\right]q
\end{equation}
where the quantity $\sqrt(qT)$ models the change in acceleration over
the sampling interval \cite{Barshalom2}. Alternatively the Singer
model in section \ref{singer} could be used.  For a manoeuvre detector
with (effective) window length $s$ declaring a manoeuvre at time $k$,
the augmented model must be initialised at time $k-s$, that is, {\em
retrospectively}. In \cite{Farooq2} it is pointed out that, strictly
speaking, the variable-dimension filter is not a filter due to the
latter requirement. A Monte Carlo comparison of the variable-dimension
and input estimation approaches appears in \cite{Barshalom2}: the
variable-dimension filter performed better on average than input
estimation, despite the latter being considerably more complex; a
two-level white noise model, the simplest algorithm of the three,
performed better than the input estimation technique and only slightly
worse than the variable-dimension filter.

\subsection{Optimal and Approximate Filters for Jump-Linear Models}\label{opt2}
In contrast to the multiple model approaches based on multi-level noise
or Markov inputs, the focus of this section is on variable structure
systems of the type (\ref{M4}), repeated here:
\begin{eqnarray}
x(k+1)&=&F(k;M(k)) x(k)+ v(k;M(k)),~v(k)\sim N\{0,Q(M(k))\}\label{M44}\\
y(k)&=&H(k;M(k)) x(k) + w(k;M(k)),~w(k)\sim N\{0,R(M(k))\}.\nonumber
\end{eqnarray}
The ``mode process'' $M(k)$ is a finite-state Markov chain taking
values in $\{M_1,\ldots,M_r\}$ with {\em known} transition probabilities
$p_{ij}$. The notation $M_i(k)$ signifies that model $M(k)=M_i$
is assumed to be in effect during the time interval $[k-1,k)$.
We also denote by  $M_l^k$ the $l$th mode history, in other words
\[
M_l^k=\{M_{l_1}(1),\ldots,M_{l_k}(k)\}=\{M_s^{k-1},M_{l}(k)\}.
\]
There are a total of $r^k$ possible mode histories at time $k$.
We can derive the optimal Bayesian (MMSE) state estimator for (\ref{M44})
by decomposing the posterior PDF of the state as follows:
\begin{equation}\label{pxrk}
{\rm p}(x(k)|Y^k)=\sum_{l=1}^{r^k}\mu_l(k)\,{\rm p}(x(k)|M_l^k,Y^k)
\end{equation}
where  $\mu_l(k)=\Pr(M_l^k|Y^k)$ are the {\em mode probabilities}.
A recursion on the mode probabilities is obtained in a similar
manner to section \ref{opt1} as
\begin{eqnarray*}
\mu_l(k)&=&\delta^{-1} {\rm p}(y(k)|M_l^k,Y^{k-1})\Pr(M_l^k|Y^{k-1})\\
&=&\delta^{-1}{\rm p}(y(k)|M_l^k,Y^{k-1})\Pr(M_l(k)|M_s^{k-1},Y^{k-1})
\Pr(M_s^{k-1}|Y^{k-1})\\
&=&\delta^{-1}{\rm p}(y(k)|M_l^k,Y^{k-1})\,p_{ij}\,\mu_s(k-1).
\end{eqnarray*}
Note that each  term ${\rm p}(y(k)|M_l^k,Y^{k-1})$ involves running a
Kalman filter on the entire mode history up to time $k$.

Computation of (\ref{pxrk}) is clearly infeasible since it involves a
sum with an exponentially growing number of terms. It is therefore
necessary to consider sub-optimal ``N-scan-back'' approaches that
``combine'' mode histories prior to time $k-N$ \cite{Barshalom7}. This
amounts to approximating the mixture PDF of the state (or measurement)
by a single Gaussian, \ie
\[
{\rm p}(x(k-N)|Y^{k-N})\sim N\{\xhat,P\}.
\]
The resulting algorithm is called a generalised pseudo-Bayesian filter
of order $N$, GPB(N), and requires only $r^N$ parallel Kalman filter
computations at each time, rather than $r^k$. In particular, for $N=1$
we obtain the GPB(1) filter which summarises the mixture PDF
(\ref{pxrk}) to time $k-1$ by a single Gaussian density and calculates
$r$ KF estimates at time $k$, collapsing this $r$-term mixture to a
single Gaussian PDF.  The probabilistic data association filter is
analogous to the GPB(1) filter with the replacement of mode hypotheses by
target/clutter associations.

The GPB(2) filter would consider the $r^2$ possible Kalman filter
estimates based on the last two manoeuvre modes $\{M(k-1),M(k)\}$
starting with a prior Gaussian density at time $k-2$. It then
recondenses the resulting mixture of $r^2$ terms to a single Gaussian
density at time $k$. In other words, the GPB(2) filter carries the
state estimate and covariance at time $k-2$ and the $r$
mode-conditioned state estimates and their covariances at time $k-1$,
running each of the $r$ conditional estimates through $r$
mode-matched Kalman filters at time $k$.  Obviously this procedure
quickly becomes unworkable, although it is generally accepted that
near-optimal performance is obtained, at least in good SNR conditions,
with GPB(2).

Instead of combining the $r^2$ mode-conditioned estimates at the end
of the filter cycle, as in GPB(2), it is possible to run only $r$ Kalman
filters in parallel, each with an appropriately weighted combination
of state estimates as a ``mixed initial condition. This reduces the
amount of processing almost to that of the GPB(1) approach while obtaining
performance near that of the GPB(2) approach.  This idea is the basis
of the interacting multiple-model (IMM) algorithm \cite{Blom}.
Returning to the mixture density (\ref{pxrk}), we have
\[
{\rm p}(x(k)|Y^k)=\sum_{j=1}^r\mu_j(k)\,{\rm p}(x(k)|M_j(k),y(k),Y^{k-1}).
\]
Now, by definition
\[
{\rm p}(x(k)|M_j(k),y(k),Y^{k-1})=
\frac{{\rm p}(y(k)|M_j(k),x(k))}
{{\rm p}(y(k)|M_j(k),Y^{k-1})}\,
{\rm p}(x(k)|M_j(k),Y^{k-1}).
\]
Conditioning on $M_i(k-1)$, it follows that
\begin{eqnarray}
{\rm p}(x(k)|M_j(k),Y^{k-1})&=&\sum_{i=1}^r \Pr(M_i(k-1)|M_j(k),Y^{k-1})
\,{\rm p}(x(k)|M_j(k),M_i(k-1),Y^{k-1})\label{mixt1}\\
&\approx&\sum_{i=1}^r \mu_{i|j}(k-1|k-1)\,
{\rm p}(x(k)|M_j(k),M_i(k-1),\xhat_i(k-1|k-1),
P_i(k-1|k-1))\nonumber
\end{eqnarray}
where $\mu_{i|j}(k-1|k-1)\keiko\Pr(M_i(k-1)|M_j(k),Y^{k-1})$
and the approximation rests on taking the mode-conditioned
Kalman filter estimates $\{\xhat_i(k-1|k-1),P_i(k-1|k-1)\}$ $l=1,\ldots,r$
as sufficient statistics for the data $Y^{k-1}$.
The mixture PDF in (\ref{mixt1}) is then approximated by a Gaussian mixture
\begin{equation}\label{mixt2}
{\rm p}(x(k)|M_j(k),Y^{k-1})=\sum_{i=1}^r\mu_{i|j}(k-1|k-1)\,
N\{x(k);\xhat_i^j(k-1|k-1),P_i^j(k-1|k-1)\}
\end{equation}
where
\begin{eqnarray*}
\xhat_i^j(k-1|k-1)&=&{\rm E}\{x(k)|M_j(k),\xhat_i(k-1|k-1),P_i(k-1|k-1)\}\\
P_i^j(k-1|k-1)&=&{\rm Cov}\{x(k)|M_j(k),\xhat_i(k-1|k-1),P_i(k-1|k-1)\}
\end{eqnarray*}
are the $r^2$ Kalman filter estimates calculated in the GPB(2) approach
(we will see that these are not required in the IMM algorithm).
The mean of the Gaussian mixture in (\ref{mixt2}) is
\begin{eqnarray*}
\xhat_j^0(k-1|k-1)&=&\sum_{i=1}^r\mu_{i|j}(k-1|k-1)\,
{\rm E}\{x(k)|M_j(k),\xhat_i(k-1|k-1),P_i(k-1|k-1)\}\\
&=&{\rm E}\{x(k)|M_j(k),\sum_{i=1}^r\mu_{i|j}\xhat_i,\,
{\rm cov}\{\sum_{i=1}^r\mu_{i|j}\xhat_i\}\}\\
&=&\sum_{i=1}^r\mu_{i|j}(k-1|k-1)\,\xhat_i(k-1|k-1)
\end{eqnarray*}
by linearity of the Kalman filter (\ie, summing many KF estimates
is the same as applying a single KF to a sum of estimates).
The covariance of (\ref{mixt2}) is
\begin{eqnarray*}
P_j^0(k-1|k-1)&=&\sum_{i=1}^r\mu_{i|j}(k-1|k-1)\,
\{P_i(k-1|k-1)+\xhat_i(k-1|k-1)\,\xhat_i(k-1|k-1)'\}\\
&~&~-\xhat_j^0(k-1|k-1)\xhat_j^0(k-1|k-1)'
\end{eqnarray*}
Only the first two moments $\xhat_j^0(k-1|k-1)$ and $P_j^0(k-1|k-1)$,
$j=1,\ldots,r$,
of the mixture (\ref{mixt2}) are retained in the IMM.

The mean and covariance of the density in (\ref{pxrk}) can then
be computed by $r$ Kalman filters each tuned to a different
{\em mixed} initial condition. Each KF in the bank of $r$
produces its estimates via
\[
\{\xhat_j^0(k-1|k-1),~P_j^0(k-1|k-1)\} \rightarrow
\mbox{\framebox[3cm][c]{Kalman Filter}}
\rightarrow
\{\xhat_j(k|k),~P_j(k|k)\}.
\]
The conditional state estimates are combined in the usual way to give
the overall state estimate and covariance
\begin{eqnarray*}
\xhat(k|k)&=&\sum_{i=1}^r\mu_j(k)\xhat_j(k|k)\\
P(k|k)&=&\sum_{i=1}^r\mu_j(k)\{P_j(k|k)+\xhat_j(k|k)\xhat_j(k|k)'\}
-\xhat(k|k)\xhat(k|k)'.
\end{eqnarray*}
Note that the $r$ estimates and covariances $\xhat_j(k|k)$ and
$P_j(k|k)$ must be stored as they are needed in the next iteration of
the algorithm. To complete the description of the IMM algorithm, we need to
develop a recursion for the mixing probabilities and evaluate
the mode probabilities $\mu_j(k)$. The mixing probabilities are computed 
with an application of Bayes' rule as
\[
\mu_{i|j}(k-1|k-1)=c_j^{-1} p_{ij} \,\mu_i(k-1),~i,j=1,\ldots,r
\]
where $c_j$ is a normalisation constant given by
\[
c_j=\sum_{i=1}^r p_{ij} \,\mu_i(k-1).
\]
The mode probabilities are computed
as in section \ref{opt1}
\[
\mu_j(k)=c^{-1} \Lambda_j(k) \sum_{i=1}^r p_{ij} \,\mu_i(k-1),~j=1,\ldots,r,
\]
in which the likelihood function for mode $j$ is evaluated as
\[
\Lambda_j(k)={\rm p}(y(k)|M_j(k),Y^{k-1})
=N\{y(k);\yhat^j(k|k-1),S^j(k)\}
\]
where the predictions for mode $j$ are based on mixed initial
conditions, {\em i.e.},
\begin{eqnarray*}
\yhat^j(k|k-1)&=&HF\xhat_j^0(k-1|k-1)\\
S^j(k)&=&HP^j(k|k-1)H'+R\\
P^j(k|k-1)&=&F P_j^0(k-1|k-1) F' + Q.
\end{eqnarray*}
The mode probabilities $\mu_j(k)$ must also be stored during the processing.
An attractive property of the IMM algorithm for manoeuvring target
tracking is that it does not require an explicit manoeuvre detector
since a range of possible manoeuvres are built into the model and these
are selected by a probabilistic weighting.

In \cite{Barshalom12,Barshalom7} a two-model and a three-model IMM
filter are compared with the standard input estimation method and the
variable-dimension filter for tracking a target executing a 90 degree
turn. The models were taken to be: $M_1$ constant-velocity, $M_2$
Wiener process acceleration model (\ref{wiener}) with acceleration
increment $q=0.001$, $M_3$ constant acceleration model with zero
process noise.  A Monte Carlo simulation showed a substantial
reduction in position error for both IMM filters during the turn, with
a factor of two decrease in RMS position error during
constant-velocity periods.  The complexity of the input estimation
filter was around 10 times that of the three-model IMM filter. It is
claimed that the results are not very sensitive to the choice of
Markov transition probabilities $p_{ij}$ as long as the matrix remains
diagonally dominant. Adjustment of the (diagonal) transition
probabilities merely trades off the peak error during the manoeuvre
with the RMS error for constant-velocity tracking.  However in
\cite{Dufour} it is shown that performance of the IMM algorithm may
depend strongly on the choice of transition probabilities for
discrimination of the correct target manoeuvring regime. It is also
indicated that careful selection of the different manoeuvre models is
required.

It is clear that the choice of models in the IMM is an important
implementational consideration. In particular the ``dynamic range'' of
the set of models should be sufficient to give an adequate description
of the anticipated target dynamics.  In \cite{Barshalom13}, a method
for implementing IMM filters with variable structure was proposed. The
algorithm adapts the models used in an IMM filter over a discrete
set. This avoids over-parametrising the dynamics (which can be as bad
as having too few models) while retaining sufficient coverage for good
tracking.  The case of a continuum of models was treated in
\cite{Maybeck1,Sheldon}.  In \cite{Gauvrit} the need for selecting the
best model set was also mentioned, although the solution presented there was
iterative rather than adaptive.

A comparison of the IMM algorithm and a Viterbi algorithm
(VA) (applied to solve the manoeuvre association problem) for a system
with multi-level white process noise was covered
in \cite{Averbuch}. Their conclusions were: 1) the performance of
both the IMM and the VA depends monotonically on the (maximal)
difference between the discrete accelerations $u_i$ and on the sampling
time, and inversely on the measurement noise; 2) the VA is
generally outperforms the IMM algorithm for high sampling rates, low
measurement noise and a large number of discrete accelerations; but 3)
the IMM provides better state estimation immediately following the
onset of the manoeuvre.

\subsection{Two-Stage Kalman Filter}
A well-known application of Kalman filtering is the simultaneous
estimation of a dynamic process with a constant, or slowly-varying
unknown bias that is correlated with the state. The dynamical state is
augmented by the constant and the estimation is performed in the usual
way. This technique has been applied to estimate the acceleration of a
manoeuvring target by treating the target acceleration as a bias term
on a constant-velocity model \cite{Alouani}. We now summarise the
approach.

Consider the following system description
\begin{eqnarray}
x(k+1)&=&F x(k)+ G u(k) + v(k)\label{Mb}\\ u(k+1)&=&u(k)+n(k)\\
y(k)&=&H x(k) + w(k)
\end{eqnarray}
where $v(k)$, $n(k)$, $w(k)$ are zero-mean, white Gaussian noise sequences
with covariance $Q_v$, $Q_n$ and $R$ respectively, $v(k)$ and $w(k)$
are uncorrelated, and $v(k)$ and $n(k)$ are jointly Gaussian with covariance
${\rm E}\{v(k)n'(k)\}=Q_{vn}$. Another way of writing (\ref{Mb}) is
\begin{eqnarray}
\left[
\begin{array}{c}
x(k+1)\\u(k+1)
\end{array}
\right]&=&\left[
\begin{array}{cc}
F&G\\0&I
\end{array}
\right]\left[
\begin{array}{c}
x(k)\\u(k)
\end{array}
\right]+\left[
\begin{array}{c}
v(k)\\n(k)
\end{array}
\right]\label{Mb2}\\
y(k)&=&\left[
\begin{array}{cc}
H&0
\end{array}
\right]
\left[
\begin{array}{c}
x(k)\\u(k)
\end{array}
\right]+w(k)\nonumber
\end{eqnarray}
Under certain observability/controllability conditions on the system
matrices \cite{Anderson}, the system (\ref{Mb2}) is stable
in the sense that the covariance of the Kalman filter converges to a
unique positive-definite solution.
From a practical standpoint, it may be undesirable to increase the
dimension of the Kalman filter. However, it is also possible to write
the recursions implied in (\ref{Mb2}) as a two-stage system: the first
stage is a Kalman filter for the system (\ref{Mb}) assuming zero bias
($u(k)\equiv 0$), the second stage is a filter based only on the bias
term $u(k)$. The outputs of the stages are combined to yield the same
estimate sequence as that which would be obtained by an augmented
filter for (\ref{Mb2}).

The bias-free filter produces $\xhat_0(k|k)$ according to the recursions below:
\begin{eqnarray*}
\xhat_0(k|k-1)&=&F\xhat_0(k-1|k-1)\\
\xhat_0(k|k)&=&\xhat_0(k|k-1)+W_0(k)\nu_0(k)\\
\nu_0(k)&=&y(k)-H\xhat_0(k|k-1)\\
P_0(k|k-1)&=&FP_0(k-1|k-1)F'+Q_0\\
P_0(k|k)&=&(I-W_0(k)H)P_0(k|k-1)\\
W_0(k)&=&P_0(k|k-1)H'S_0^{-1}(k)\\
S_0(k)&=&HP_0(k|k-1)H'+R,
\end{eqnarray*}
for some covariance $Q_0$ yet to be defined.
The bias filter computes its estimate via:
\begin{eqnarray*}
\uhat(k|k-1)&=&\uhat(k-1|k-1)\\
\uhat(k|k)&=&\uhat(k|k-1)+W_u(k)\nu_0(k)-C(k)\uhat(k|k-1)\\
P_u(k|k-1)&=&P_u(k-1|k-1)+Q_u\\
P_u(k|k)&=&(I-W_u(k)C(k))P_u(k|k-1)\\
W_u(k)&=&P_u(k|k-1)C(k)'S_u^{-1}(k)\\
S_u(k)&=&C(k)P_u(k|k-1)C(k)'+HP_0(k|k-1)H'+R,
\end{eqnarray*}
where the bias gain is computed using
\begin{eqnarray*}
C(k)&=&HU(k)\\
U(k)&=&FV(k-1)+G=F(I-W_0(k-1)H)U(k-1)+G\\
V(k)&=&(I-W_0(k)H)U(k).
\end{eqnarray*}
The two-stage Kalman filter estimate is formed using the bias-free state
and bias estimates 
\begin{eqnarray*}
\xhat(k|k)&=&\xhat_0(k|k)+V(k)\uhat(k|k)\\
\xhat(k|k-1)&=&\xhat_0(k|k-1)+U(k)\uhat(k|k-1)\\
P_{11}(k|k)&=&P_0(k|k)+V(k)P_u(k|k)V(k)'\\
P_{12}(k|k)&=&V(k)P_u(k|k)\\
P_{12}(k|k-1)&=&U(k)P_u(k|k-1)\\
P_{22}(k|k)&=&P_u(k|k)\\
P_{22}(k|k-1)&=&P_u(k|k-1)
\end{eqnarray*}
where $P_{ij}$ are the blocks of the covariance matrix for the
augmented state $(x',u')'$.  A theorem in \cite{Alouani} shows that
the two-stage Kalman filter is equivalent to the augmented Kalman
filter when
\begin{eqnarray}
Q_{vn}&=&U(k+1)Q_u\nonumber\\
Q_0&=&Q_v-U(k+1)Q_nU'(k+1)\geq 0.\label{qq}
\end{eqnarray}
The positive semi-definiteness of $Q_0$ is required for stability
of the filter.

The overall complexity of the bias-free and bias filters is less than
that of the augmented Kalman filter in most cases of interest. The
operation of the two-stage filter is akin to a variable-dimension
filter with a higher-order (acceleration) estimator that can be
switched off in quiescent mode. A standard exponentially-weighted
innovations manoeuvre detector (section \ref{detec}) is used to detect
the onset of the manoeuvre.  If the effective window length is
$\Delta$, and a manoeuvre is detected at time $k$, the acceleration
(bias) filter, but not the bias-free filter, must be initialised
(retrospectively) at time $k-\Delta$ to correct the constant-velocity
filter estimates.  The covariance $Q_0$ is also increased during the
manoeuvre. Termination of the manoeuvre is sensed by monitoring the
innovations of the bias-free filter. In a simulation in \cite{Alouani}
the two-stage Kalman filter was compared with the variable-dimension
filter. The results indicated that the estimation errors were similar
before and during a manoeuvre, but the two-stage filter detected the
end of the manoeuvre more rapidly.

Regarding initialisation of the two-stage Kalman filter, the initial
bias estimate is taken to be uncorrelated with the initial bias-free
state estimate
\[
P_{12}(0|0)=0\implies V(0)=0,~U(0)=0.
\]
The bias-free process noise $Q_0$ (\ref{qq}) sets the desired response of the
constant-velocity filter; the bias filter process noise $Q_u$ determines
the manoeuvre response.

The two-stage Kalman filter has been applied within the IMM framework
by \cite{Munir,Munir1} with the objective of making the assumed target
acceleration a variable parameter in the IMM. In this way, only two
filters per co-ordinate, rather than several
fixed-acceleration models, are required to cover the possible range of target
motions. The results of this study indicate similar performance at
lower complexity compared with the standard IMM approach. A difficulty
is the delay required by the two-stage Kalman filter in estimating the
manoeuvre acceleration, although this is somewhat alleviated by the
subsequent matching of the IMM model to the target acceleration.

In \cite{Blair3} an interacting multiple model algorithm with bias
estimation was proposed, again based on a two-stage Kalman filter.
By using only bias filters in the IMM as opposed to augmented-state
Kalman filters, a complexity reduction is achieved. A simulation
was provided for an IMM with 3 bias models corresponding to:
zero bias (constant velocity), constant non-zero bias (constant
acceleration), and constant bias with a constant-speed constraint.
The performance obtained was similar to a 3-model IMM filter
while requiring roughly 50\% of the computation.

\begin{figure}[h]
\centerline{\psfig{figure=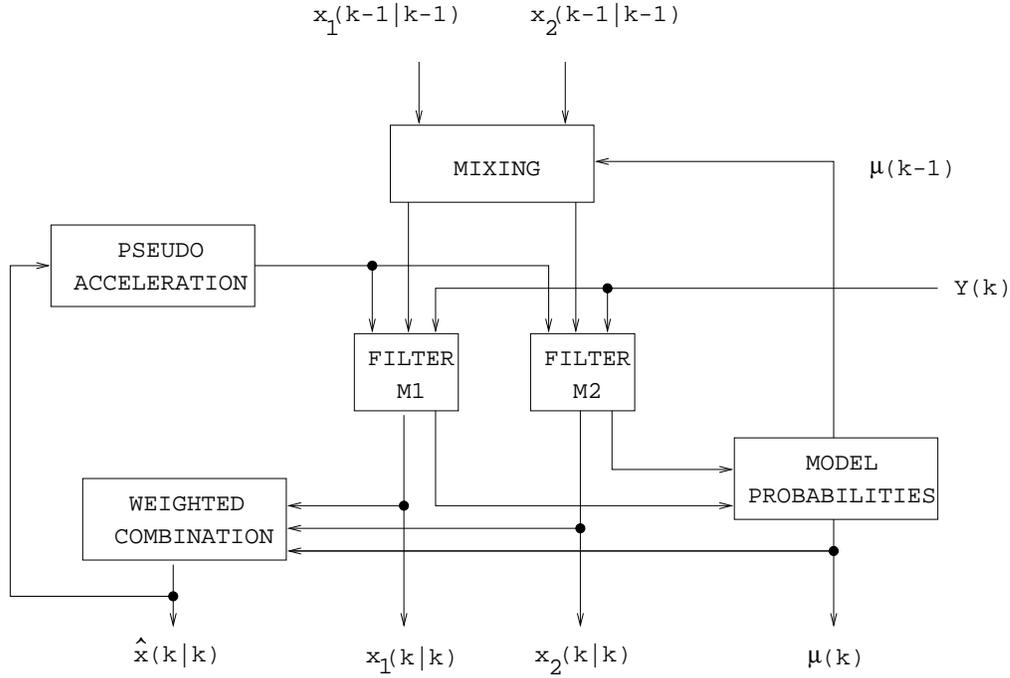,height=9cm}}
\caption{\protect{\small Block diagram of the interacting
multiple model algorithm with acceleration 
compensation.}}
\label{Fig5}
\end{figure}

Lastly an interacting acceleration compensation (IAC) technique was
put forward in \cite{Watson3} as a means of obviating the requirement
for manoeuvre detection of the two-stage Kalman filter.  The IAC
technique, illustrated in Fig. \ref{Fig5} (from \cite{Blair1}),
assumes a two-stage estimator with a bias-free (constant-velocity)
filter and two acceleration (bias) models: one for a constant-velocity
target and one for constant acceleration.  Since the IMM does not
employ a conventional manoeuvre detector, it can provide an
acceleration estimate in real time with which to compensate the
bias-free filter. The IAC algorithm operates at roughly half the
computational cost of an IMM filter with two models, but provides
similar performance.

\subsection{IMM Algorithm for Fixed Interval Smoothing}
A fixed-interval smoothing approach to manoeuvring target tracking has
been derived in \cite{Helmick}, following earlier work by
\cite{Fraser,Blom1}. The forward-running and backward-running Kalman filters
in the usual fixed-interval smoother are replaced by interacting
multiple model filters, together with a technique for fusing the
conditional estimates from the two filters. The backward-running
filters are obtained by time-reversal of the corresponding
forward-running filters. Although the proposed algorithm is a batch
algorithm with high complexity, its high performance makes it useful
as an estimator for generating benchmark tracks in ground
truthing or for sensor alignment in multi-sensor tracking systems.

In the notation of section \ref{opt2}, assuming $r$ models, suppose
that at time $k$ model $M_i(k)$ is in effect for the forward filter
with the backward filter assuming model $M_j(k+1)$. The conditional
state ``predictions'' and their covariances are computed as
$\xhat_i^F(k|k-1)$, $P_i^F(k|k-1)$, $\xhat_i^B(k|k-1)$,
$P_i^B(k|k-1)$. Assuming a fixed interval of size $N$, the predictions
are used to obtain $r^2$ fused initial estimates according to
\begin{eqnarray*}
\xhat_{ij}^0(k|N)&=&P_{ij}^0(k|N)[
(P_i^F(k|k-1))^{-1}\xhat_i^F(k|k-1)+(P_j^B(k|k-1))^{-1}\xhat_j^B(k|k-1)]\\
P_{ij}^0(k|N)&=&(P_i^F(k|k-1))^{-1}+(P_j^B(k|k-1))^{-1}.
\end{eqnarray*}
These quantities are then mixed to yield prior state and covariances
for the Kalman smoother matched to model $M_j(k)$
\begin{eqnarray*}
x_i^0(k|N)&=&\sum_{j=1}^r \mu_{j|i}(k+1|N)\,\xhat_{ji}^0(k|N)\\
P_i^0(k|N)&=&\sum_{j=1}^r \mu_{j|i}(k+1|N)\,[P_{ji}^0(k|N)+\xhat_{ij}^0(k|N)
\xhat_{ij}^0(k|N)']-x_i^0(k|N)x_i^0(k|N)'.
\end{eqnarray*}
The mode probabilities $\mu_j(k|N)$ are updated via the smoothed
measurements $\yhat_j(k|N)$ from smoother $M_j(k)$. Finally the
conditional estimates are combined to give
\[
\xhat(k|N)=\sum_{j=1}^r\mu_j(k|N)\,\xhat_j(k|N)
\]
The IMM smoother was compared under simulation in \cite{Helmick}
with a single model (constant-velocity) smoother and yielded similar
performance, although the former method seemed to produce larger spikes
at the manoeuvre onset times.

\subsection{Expectation-Maximisation Algorithm}
This technique, from \cite{Pulford10}, utilises multi-level process
noise of the form (\ref{M3}) to model the unknown target
accelerations. A zero acceleration level is included for a
constant-velocity target.  We give only a brief description of the
technique here.

A length $K$ batch of uncluttered target measurements is assumed
available for processing.  The expectation maximisation (EM) algorithm
\cite{Dempster,Baum2,Moon} is applied to the problem of
estimating the MAP sequence of target accelerations $U^K$:
\begin{equation}\label{em}
\arg\max_{U^K} ~
{\rm p}(X^K,Y^K,U^K)
\end{equation}
where $X^K$ denotes the (unknown) state sequence of length $K$ and $Y^K$
the measurement sequence.
This results in a multi-pass, batch estimator of $U^K$.  At each pass
an objective function is evaluated (the E step) and then maximised
(the M step).  The expectation step involves computation
of state estimates from a bank of Kalman smoothers tuned to the possible
manoeuvre sequences.  The maximisation step is efficiently
implemented using forward dynamic programming.

An on-line estimator is also derived using a modified EM-type cost
function. To obtain a dynamic programming recursion, the target state
is assumed to satisfy a Markov property with respect to the manoeuvre
sequence. This results in a recursive but sub-optimal estimator
implementable on a Viterbi trellis. The transition costs of the latter
algorithm, which depend on filtered estimates of the state, are
compared with the costs arising in a Viterbi-based manoeuvre estimator
developed in \cite{Averbuch}. It is shown that the two criteria
differ only in the weighting matrix of the quadratic part of the cost
function. 

Testing of the batch EM manoeuvre tracking algorithm (reported in
\cite{Pulford10a}) has demonstrated superior transient error
performance during manoeuvres compared with the recursive version,
which has errors of similar magnitude to the approach in
\cite{Averbuch}.  The false manoeuvre alarm performance of the latter
algorithm was the best of the three techniques tested in
\cite{Pulford10a}.

\section{Manoeuvring Target Tracking in Clutter}\label{clutter}
\begin{figure}[h]
\centerline{\psfig{figure=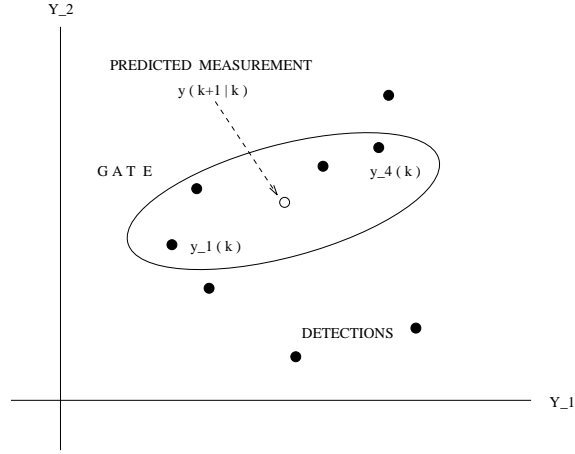,height=6cm}}
\caption{\protect{\small Gating is applied to measurements
to reduce the processing required for tracking in a cluttered
environment. The gate is centred on the predicted measurement.}}
\label{Fig6}
\end{figure}

Tracking a constant-velocity target in the presence of false
measurements or clutter involves solving the {\em data association}
problem. This is the problem of deciding which, if any, of a number of
candidate detections is more likely to have arisen from the target of
interest. The data association problem involves prediction of the
measurement to the next scan followed by a process that correlates
measurements with the prediction. In heavily cluttered environments it
is well to apply {\em gating} to reduce the number of plausible target
measurements to a reasonable number for processing.  The process of
gating or measurement validation is covered in
\cite{Farina,Blackman,Barshalom2} and illustrated in Fig. \ref{Fig6}.

\begin{figure}[h]
\centerline{\psfig{figure=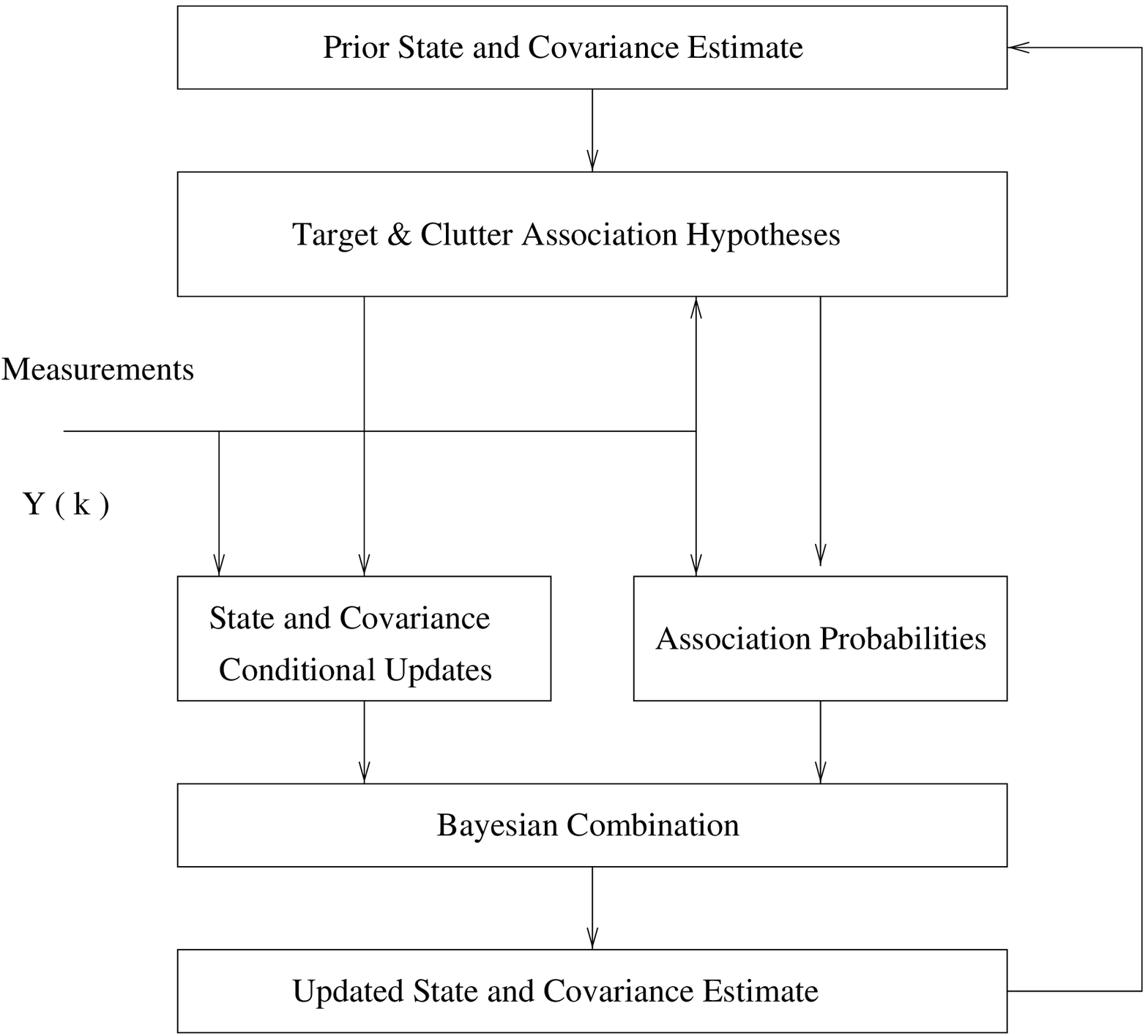,height=9cm}}
\caption{\protect{\small Block diagram of a probabilistic
data association filter. The same processing applies to any algorithm
that approximates the measurement density by a single Gaussian
at each scan.}}
\label{Fig7}
\end{figure}

Since the optimal (Bayesian) solution \cite{Singer} to the data
association problem has combinatorially increasing computational
requirements, there are many sub-optimal approaches of which the
nearest neighbour Kalman filter, probabilistic data association (PDA)
\cite{Barshalom1}, the track-splitting filter \cite{Smith}, and the
multiple-hypothesis (MHT) filter \cite{Reid} are among the more
common.  A block diagram of the PDA filter is given in
Fig. \ref{Fig7}, although the same structure applies to any
single-scan Bayesian algorithm for state estimation.

As we mentioned in the Introduction, the problem of tracking a
manoeuvring target is fundamentally at odds with the data association
problem. This is because any method that employs distance from a
predicted measurement (\ie, the filter innovation) to determine the
onset of a manoeuvre may be falsely triggered in the presence of
nearby clutter detections.  In this section we review the optimal
Bayesian approach to tracking a manoeuvring target in clutter and its
natural sub-optimal realisations. We also outline some adaptive
PDA-based schemes (variable-dimension PDA and IMM-PDA). We also
describe some approaches that take advantage of signal amplitude
distributions to improve the discrimination of target from clutter.
These approaches can be integrated at low cost into most existing
tracking filters.  The discrimination of false alarms from target
measurements can also be aided by consistency testing of velocity and/or
and acceleration estimates in addition to the filter innovation.

\subsection{Optimal Bayesian Filter for Manoeuvring Target in Clutter}
We present a derivation of the optimal Bayesian filter (OBF)
\cite{Kenefic} in this part.  The manoeuvring target model (\ref{M3})
presented earlier is modified for the case of cluttered measurements
as
\begin{eqnarray}
x(k+1)&=&\left\{
\begin{array}{cc}
F(k) x(k)+G(k)u_1(k)+v(k)&\mbox{for model 1}\\
F(k) x(k)+G(k)u_2(k)+v(k)&\mbox{for model 2}\\
\vdots&\vdots\\
F(k) x(k)+G(k)u_r(k)+v(k)&\mbox{for model r}
\end{array}
\right.
\nonumber\\
y(k)&=&\left\{
\begin{array}{ll}
H(k) x(k) + w(k)&\mbox{for the target}\\
\mbox{clutter}&\mbox{otherwise}.
\end{array}
\right.
\label{M3c}
\end{eqnarray}
The target accelerations $u(k)$ are a finite-state Markov chain with
known transition probabilities $p_{ij}$, with $u_i(k)$ denoting the
event $\{u(k)=u_i\}$, $i=1,\ldots,n$. The usual notation for cluttered
measurement sets is adopted: $Y(k)=\{y_1(k),\ldots,y_{m_k}(k)\}$ (at
scan $k$), $Y^k=\{Y(1),\ldots,Y(k)\}$ (measurement set up to time
$k$). The {\em association event} $\theta_i(k)$ signifies the event
that measurement $y_i(k)$ is due to the target ($i=1,\ldots,m_k$), or
no measurement is from the target ($i=0$).

Seeking to evaluate the probability of a particular joint measurement
association and manoeuvre sequence, we define the $l$th possible
measurement history (recursively) as
\[
\Theta_l^k=\{\Theta_s^{k-1},\theta_{i_l}(k)\},~l=1,\ldots,L_k
\]
where $L_k=(1+m_1)(\cdots)(1+m_k)$,
and the $p$th possible manoeuvre history as
\[
\Phi_p^k=\{\Phi_q^{k-1},u_{i_p}(k)\},~p=1,\ldots,n^k.
\]
The set of joint measurement and manoeuvre association events
is $\chi_{lp}^k=\{\Theta_l^k,\Phi_l^k\}$ in which there are
$L_k\,n^k$ elements!

Writing $\chi_{lp}^k$ as $\{\chi_{sq}^{k-1},\chi_{lp}(k)\}$
where $\chi_{lp}(k)=\{\theta_{i_l}(k),u_{i_p}(k)\}$, we obtain
the conditional mean estimator for $x(k)$ as
\begin{equation}\label{opt3}
\xhat(k|k)\keiko {\rm E}\{x(k)|Y^k\}=
\sum_{p=1}^{n^k}
\sum_{p=1}^{L_k}
\beta_{lp}(k)\,\xhat_{lp}(k|k)
\end{equation}
where the conditional state estimates are defined as $\xhat_{lp}(k|k)=
{\rm E}\{x(k)|\chi_{lp}^k\}$. The association probabilities
$\beta_{lp}(k)=\Pr(\chi_{lp}^k|Y^k)$ satisfy a recursion, namely
\[
\beta_{lp}(k)=c_k^{-1}{\rm p}(Y(k)|\chi_{sq}^{k-1},\chi_{lp}(k),Y^{k-1})
\Pr(\chi_{lp}(k)|\chi_{sq}^{k-1},Y^{k-1})\beta_{sq}(k-1).
\]
in which $c_k$ is a normalisation factor.
Furthermore the terms on the right hand side are evaluated as
\[
{\rm p}(Y(k)|\chi_{sq}^{k-1},\chi_{lp}(k),Y^{k-1})=\left\{
\begin{array}{ll}
V_G^{-m_k}&i_l=0\\
V_G^{-m_k+1}P_G^{-1} N\{y_{i_l}(k);\yhat_{sq}(k|k-1),S_{sq}(k)\}&i_l>0,
\end{array}
\right.
\]
where $\yhat_{sq}(k|k-1)={\rm E}\{y(k)|\chi_{sq}^{k-1},Y^{k-1}\}$ and
its covariance $S_{sq}(k)$ are computed by Kalman filtering the $s$th
measurement sequence assuming the $q$th manoeuvre history, and $V_G$
is the volume of the validation gate at time $k$.  The computation of
the association probabilities is completed with
\[
\Pr(\chi_{lp}(k)|\chi_{sq}^{k-1},Y^{k-1})=
\Pr(u(k)=u_{i_p}|u(k-1)=u_{i_q})\,\Pr(\theta_{i_l}(k)|Y^{k-1})
\]
and
\[
\Pr(\theta_{i_l}(k)|Y^{k-1})=\left\{
\begin{array}{ll}
c^{-1}\frac{1}{m_k}\,P_D\,P_G\,g_c(m_k-1)&i_l=1,\ldots,m_k\\
c^{-1}(1-P_DP_G)\,g_c(m_k)&i_l=0,
\end{array}
\right.
\]
where $P_D$ is the target detection probability, $P_G$ the gate
probability and it is customary to assume that the number of clutter
points in the gate is given by a Poisson distribution
\[
g_c(m)=\exp(-\lambda V_G)\frac{(\lambda V_G)^m}{m!},~m=0,1,\ldots
\]
The state error covariance is computed as
\[
P(k|k)=\sum_{p=1}^{n^k}
\sum_{p=1}^{L_k}
\beta_{lp}(k)\,\xhat_{lp}(k|k)
\left[
P_{lp}(k|k)+\xhat_{lp}(k|k)\xhat_{lp}(k|k)'
\right]-
\xhat(k|k)\xhat(k|k)'.
\]
The optimal Bayesian filter (\ref{opt3}) is of course not realisable
due to its increasing memory and computational requirements.  There
exist $(N,M)$ scan approximations to it that lump measurement
histories identical over the previous $N$ scans and manoeuvre
histories identical over the previous $M$ scans (see \cite{Kenefic}).
These techniques amount to matching a single Gaussian density to the
mixture (\ref{opt3}) at the appropriate time. Note that the MAP state estimator
would take only the estimate $\xhat_{lp}(k|k)$ corresponding to the
maximum association probability $\beta_{lp}(k)$ at time $k$.

\subsection{Single-Scan Pseudo-Bayesian Filter}
In \cite{Gauvrit} a sub-optimal Bayesian adaptive filter for a
manoeuvring target system of the following form is derived:
\begin{eqnarray*}
x(k+1)&=&F(k) x(k)+v(k),~v(k)\sim N\{0,Q(k)\}\\
y(k)&=&\left\{
\begin{array}{lll}
H(k) x(k) + w(k)&w(k)\sim N\{0,R(k)\}&\mbox{for the target}\\
\mbox{clutter}&&\mbox{otherwise}.
\end{array}
\right.
\end{eqnarray*}
In this formulation the process and measurement noise covariances
are {\em unknown} and take values in a discrete set
\[
Q(k)\in\{Q_1,\ldots,Q_p\},~
R(k)\in\{R_1,\ldots,R_l\}.
\]
Let $\phi_{ij}(k)$ denote the event that at time $k$
$\{Q(k)=Q_i,R(k)=R_j\}$, and define the association events
$\theta_i(k)$ $i=0,1,\ldots,m_k$ as before. It follows that the
conditional mean estimator of the state is expressible as
\begin{eqnarray*}
\xhat(k|k)&=&\sum_{j=1}^p\sum_{l=1}^q
\Pr(\phi_{lj}(k)|Y^k)\sum_{i=0}^{m_k}
{\rm E}\{x(k)|\theta_i(k),\phi_{lj}(k), Y^k\}\\
&=&\sum_{j=1}^p\sum_{l=1}^q
\Pr(\phi_{lj}(k)|Y^k)\,\xhat_{jl}(k|k)
\end{eqnarray*}
where, following the usual pattern,
\[
\xhat_{jl}(k|k)=\sum_{i=0}^{m_k}\beta_{ijl}(k)\,{\rm E}\{x(k)|\theta_i(k),
\phi_{lj}(k), Y^k\},
\]
and
\[
\beta_{ijl}(k)\keiko \Pr(\theta_i(k),\phi_{lj}(k)|Y^k).
\]
This multiple-model PDA technique is also described in \cite{Barshalom9}.

The filter is made practicable by summarising the prior state density
based on $Y^{k-1}$ by a single Gaussian with mean $\xhat(k-1|k-1)$ and
covariance $P(k-1|k-1)$ (as in the GPB(1) or PDA approaches).  Despite
this single-scan approximation of the Gaussian mixture density, the
computational requirements of this adaptive PDA filter remain high. In
particular it is difficult in practice to determine a suitable grid
size for discretisation of the covariance matrices. In a bearings-only
tracking example for a ship travelling in a non-straight course with
an average 10 clutter measurements per look, \cite{Gauvrit} used $p=6$,
$q=6$ cells, or a total of 36 PDA filters.

\subsection{Joint Manoeuvre and Measurement Association}\label{jmma}
A similar approach to the preceding section was taken by
\cite{Sengupta,Kosuge} for tracking multiple manoeuvring targets in
clutter, with \cite{LeeHJ} covering in the single-target case. A multi-level
white noise acceleration model was used with the input a finite Markov
chain $u(k)\in\{u_1,\ldots,u_m\}$ with known transition probabilities
$p_{ij}$. Gating was applied to validate measurements for tracking. The filter
derivation, again based on a single-scan Gaussian approximation, is
briefly summarised below.

Let $\theta_i(k)$ denote the $i$th association event and $u_j(k)$ the
$j$th manoeuvre model. The association probabilities are computed by
invoking the independence of the measurement and manoeuvre
associations, {\em viz.}
\begin{eqnarray*}
\beta_{ij}(k)&=&\Pr(\theta_i(k),u_j(k)|Y^k)\\
&=&c^{-1} {\rm p}(Y(k)|\theta_i(k),u_j(k), Y^{k-1})
\Pr(\theta_i(k),u_j(k)|Y^{k-1}).
\end{eqnarray*}
The first term on the right is calculated as in standard PDA.
The last term in the expression is
\begin{eqnarray*}
\Pr(\theta_i(k),u_j(k)|Y^{k-1})&=&\Pr(\theta_i(k)|u_j(k),Y^{k-1})
\Pr(u_i(k)|Y^{k-1})\\
&=&\Pr(\theta_i(k)|Y^{k-1})\sum_{l=1}^m p_{lj}\Pr(u_l(k-1)|Y^{k-1})
\end{eqnarray*}
whence standard PDA-type calculations can be used to complete
the derivation. The state estimate is given by
\[
\xhat(k|k)=\sum_{i=0}^{m_k}\sum_{j=1}^m \beta_{ij}(k)\,\xhat_{ij}(k|k)
\]

The simulation results in \cite{Sengupta} assume a high probability of
detection $P_D=0.95$ and a very low clutter density of 0.1--0.2
returns in the gate on average. For this low clutter case, the
technique appears to be able to track multiple manoeuvring targets
that overlap, although the estimation accuracy is degraded during
manoeuvres. The results in
\cite{Kosuge}, which assume $P_D=1$ and $P_{FA}=0.1$ and 6 possible
target accelerations, are hard to interpret. It is also unclear how
the probability of false alarm indicated in the simulations translates to
the usual clutter density.

\subsection{Variable-Dimension PDA}\label{vdpda}
Variable dimension PDA \cite{Birmiwal} is a direct extension of the
variable-dimension filtering approach described in sections
\ref{detec} and \ref{vardim} to the case of cluttered measurements. 
Clutter is taken into account by replacing the usual Kalman filters
for constant-velocity and constant-acceleration targets with PDA
filters, with the idea that the constant-velocity PDA filter is
maintained until a manoeuvre is detected at which time a
constant-acceleration model takes over. The difficulty in this type of
approach is the reliable detection of a manoeuvre in the presence of
false measurements. A series of statistical tests based on the
normalised filter innovations are needed to solve this problem. 
Since there is more than one measurement at each scan, there are many
different sequences of innovations.  The method is therefore to test
all possible sequences of innovations $\{\rho_i(j)\}$,
$j=k-N+1,\ldots,k$ over a window of length $N$ and initiate a new
constant-acceleration PDA filter each time a sequence is found that
satisfies the following tests.
\begin{enumerate}
\item 
Gating: $\{\rho_i(j)\}$ should be contained in a series of gates of increasing
size.
\item 
Significance of the accelerations estimate: the innovations are white and zero-mean for a non-manoeuvring target, a target acceleration corresponds to
correlation in this sequence. The presence of a non-zero acceleration
is tested at a given significance level.
\item 
Goodness of fit test for the acceleration model: the MMSE
of the residuals is computed allowing for the estimated acceleration.
\item 
If the previous tests are all satisfied, a measurement is expected
at time $k+1$ near the prediction based on the acceleration estimate.
This is used to initiate a higher order Kalman filter which
is then continued after time $k+2$ using PDA to track the manoeuvre
in clutter.
\end{enumerate}
Since an innovation sequence corresponding to false alarms
may pass all the above tests, the constant-velocity PDA filter
is continued along with any manoeuvring PDA filters. This
decreases the risk of track loss in clutter.

Track maintenance rules are provided in \cite{Birmiwal} for
terminating and merging the extra tracks that are produced by this
algorithm. The finish of the manoeuvre is detected by testing
innovations in the gate of the PDA filter assuming a zero acceleration.
Although the technique does not require retrospective initialisation,
many PDA filters are needed to track a manoeuvring target in moderate clutter.

The performance of the variable-dimension PDA tracker is reported
to be reasonable in moderate clutter densities (0.024 clutter detections
on average in a $2\sigma_x\times 2\sigma_y$ measurement noise cell).
In addition to the large number of filters that have to be maintained
to minimise track loss, there are several thresholds that need to be set
for the statistical tests described above.

\subsection{IMM-PDA}\label{immpda}
A natural extension of the IMM algorithm for tracking
a manoeuvring target in a cluttered environment appears in
\cite{Blom2,Houles}. This technique, called interacting multiple-model
probabilistic data association (IMM-PDA), replaces the standard Kalman
filters in the IMM algorithm (section \ref{opt2}) with PDA filters.
The target models can be taken as constant-velocity and
constant-acceleration for the manoeuvring target case, or
constant-velocity and ``no target'' (non-existent target) models for the
algorithm in
\cite{Barshalom4,Barshalom5} which is designed for automatic track
initiation and tracking in clutter. (The technique is clearly not
restricted to just two models.) The transitions between the models are
Markov with known transition probabilities. The major difference in
the IMM computations is that the mode probabilities are formed using
the PDA ``likelihood'' function
\begin{eqnarray*}
\Lambda_j(k)&=&{\rm p}(Y(k)|M_j(k),Y^{k-1})\\
&=&V_G^{-m_k}(1-P_DP_G)+V_G^{-m_k+1}\frac{P_D}{m_k}
N\{y_i(k);\yhat_j(k|k-1),S_j(k)\}\\
\yhat_j(k|k-1)&=&{\rm E}\{y(k)|M_j(k),Y^{k-1}\}
\end{eqnarray*}
where the symbols have usual meanings. The prediction $\yhat_j(k|k-1)$
is obtained from the IMM filter for model $M_j(k)$. The other computations
are identical to the standard IMM algorithm. The performance of the IMM-PDA
filter appears to be good for its level of complexity.

A further variation on this theme is provided in \cite{Helmick1} which
uses ``integrated PDA'' filters instead of PDA filters in the IMM-PDA
algorithm, leading to the title IMM-IPDA. The integrated PDA filter
\cite{Musicki1} incorporates a target existence model and hence
provides a track confidence measure. The target existence model is
identical to the one in \cite{Colegrove1}, although
the association probability calculations are marginally different,
replacing the usual $m_k$ term in PDA by an expected number of
measurements in the gate $\hat{m}_k$. The IMM-IPDA algorithm has
essentially the same complexity as the IMM-PDA algorithm and the track
confidence measure assists in automatic track formation and
maintenance. Simulations evidenced in \cite{Helmick1} point to
acceptable performance against constant-velocity, constant-speed
turning and linearly accelerating targets in moderate clutter. 

In \cite{Watson6} the IMM-IPDA algorithm was put forward as a solution
to the second benchmark problem for manoeuvring target tracking in
the presence of clutter and counter-measures. The algorithm also
monitors the predicted position error $\epsilon_x$ and automatically
adjusts the radar revisit time to ensure that $\epsilon_x$ does not
exceed a threshold.

An additional modification of the IMM-PDA idea appeared in \cite{Musicki3}
called integrated probabilistic data association with prediction-oriented
multiple models (IPDA-PMM). The method uses a common prediction equation
across all models to simplify the computations. Simulations
for a slowly manoeuvring target in low density clutter show modest
improvements over a constant-acceleration PDA filter.

\subsection{Multi-hypothesis Techniques}\label{mht}
Multi-hypothesis tracking (MHT) is a high performance,
``measurement-oriented'' multiple target tracking technique developed
in \cite{Reid}. Although not originally considered in the MHT
framework, more recent work on MHT has focussed on the incorporation
of manoeuvres.  A detailed description of MHT is contained in
\cite{Blackman} and we need only mention its essential features. MHT
is a Bayesian algorithm that computes state estimates for all
measurement hypotheses and their corresponding posterior
probabilities. The hypothesis set allows for a measurement to be due
to: (i) an existing target; (ii) a false alarm; or (iii) a new
target. The inclusion of the third hypothesis allows MHT to initiate
new target tracks automatically.

Due to the very high computational requirements of the MHT approach,
efficient pruning of the decision tree is needed (see references in
\cite{Koch3}). This is effected by clustering non-competing
measurements and tracks, combining tracks with similar state
estimates, deleting low-confidence tracks, pruning based on prior
information (manoeuvrability constraints), and preventing initiation
on measurements that have a high correlation to existing tracks.

The inclusion of target manoeuvres in MHT has been approached in
\cite{Koch3} via a correlated-noise acceleration (Ornstein-Uhlenbeck)
model in the context of formation tracking.  The IMM and GPB
algorithms have been applied within the MHT framework in \cite{Koch2}
to represent the PDF sum over mode histories by finite Gaussian
mixtures. A performance comparison is given for the IMM, GPB(1),
GPB(2) and a MHT-IMM filter with truncation depth of 3 scans. It is
concluded in \cite{Koch2} that the 3 scan-back approximation is
near-optimal for the benchmark tracking problem \cite{Blair5}. The
same author also applied fixed-lag Kalman smoothing or {\em
retrodiction} to improve the state estimates obtained by the MHT-IMM
algorithm and this is reported in \cite{Koch1}.  An IMM-MHT algorithm
called ``IM3HT'' has also been reported in \cite{Farina3}. The IM3HT
algorithm is a natural embedding of IMM in the MHT framework using
multiple models to cover possible target accelerations and multiple
hypotheses to allow for false alarms and multiple targets.
Comparisons of IM3HT with a MHT algorithm incorporating manoeuvre
detection logic are given for target accelerations of up to 5 g for
detection probabilities of 0.8 and 1.0.

\subsection{Filters Using Amplitude Information}
In the same way that Doppler frequency is used to enhance tracking and
track initiation performance over techniques employing only position
information, several authors have investigated the possibility of
using signal amplitude information (or power, SNR, \etc) to increase
the ability of the tracking algorithm to discriminate target
measurements from clutter.  More generally, any non-dynamical
``feature'' information (for instance target orientation, \etc) could
be used for this purpose, if it is available. The paper \cite{Dillard}
describes a distribution-free procedure for target detection in
track-while-scan and phased-array radars.  The detection procedure is
distribution-free because it uses only the ``rank'' (or ordering) of
the received measurement to determine whether a target signal is
present or absent. Thus if the rank of the measurement is $Z _{ij}$ in
range bin $i$ and beam position $j$ (and at time $k$), a signal is
declared present if $Z_{ij}\geq T$ or absent otherwise where $T$ is
the detection threshold.

The probability of false alarm for such a procedure depends only on
the number of reference cells over which the ranking takes place, and
not on the signal and noise distributions $\Pr(Z_{ij}=R|S+N)$ and 
$\Pr(Z_{ij}=R|N)$. The latter distributions are needed, however, for
computing the detection probability.

The use of ``rank'' information, as described in \cite{Dillard} was
applied to enhance the performance of the PDA filter for tracking a
manoeuvring target in clutter \cite{Nagarajan}. They also consider an
{\em ad hoc} procedure for adapting the detection threshold to trade
off computation (due to excessive clutter) and detection performance.
The only algebraic difference is in the calculation of the association
probabilities for PDA, which in this case depend on a likelihood ratio
(we will detail this modification shortly). The signal plus noise
amplitude distribution was assumed to be Rayleigh, corresponding to a
Swerling type-2 target \cite{Swerling}. In simulations at
probabilities of detection between 0.6 and 0.9, performance of the
amplitude-aided PDA filter was compared with that of a
(constant-acceleration) PDA filter for tracking a target executing a
constant-speed turn.  Apart from the obvious advantage of good quality
signal amplitude information, it is claimed that the method in
\cite{Nagarajan} can track a manoeuvring target in clutter densities
as high as 10 points (average) in the validation gate.

Similar work on the inclusion of amplitude information in PDA was
carried out in \cite{Lerro1}, and we now outline the basic approach.
The probability distributions $p_1(\cdot)$ and $p_0(\cdot)$ of the
received signal envelope $a(k)$ under the signal present
$H_1$ and signal absent $H_0$ hypotheses are assumed to be known. The
system model uses an augmented measurement vector $y_a(k)$ defined by
\[
y_a(k)=\left[
\begin{array}{c}
y(k)\\a(k)
\end{array}
\right].
\]
The state vector is not augmented, although this could be
accomplished if filtering of the signal amplitude is required.
The equations of the PDA filter are unchanged except for the computation
of the conditional measurement density, which now must allow for the
different probability densities of target and clutter measurements:
\[
{\rm p}(y_i(k)|\theta_i(k),Y^k)=\left\{
\begin{array}{cc}
P_G^{-1} N\{y_i(k);\yhat(k|k-1),S(k)\}\,p_1(a(k))&i=1,\ldots,m_k\\
V_G^{-1}\,p_0(a(k))&i=0
\end{array}
\right.
\]
This introduces a likelihood ratio into the calculation of the association
probabilities (once they are normalised). Effectiveness of the
technique relies on the accuracy with which the respective amplitude
distributions are modelled.

Signal amplitude information was also included in the IMM-PDA
algorithm in \cite{Lerro1a}. This uses the likelihood ratio arising
from the signal-plus-noise and noise-only distributions to alter the
PDA association probabilities as illustrated above. The modification
is useful for enhancing track formation and tracking of manoeuvring
targets in clutter through more effective signal discrimination. In
the same paper a track acceptance test that uses a two-dimensional
threshold, depending on both the target confidence and the probability
distribution for false track duration, is described in detail. The
target confidence measure is obtained by including a ``no target''
model in the IMM filter and the distribution of false track lifetimes
is obtained via Monte Carlo simulation. The IMM filter is implemented
with three other models: constant-velocity, left-turn and
right-turn. The performance is claimed to be superior to that of the
IMM-PDA filter (section \ref{immpda}) especially for weak targets and
in heavy clutter.

Amplitude information has also been applied in the context of
multi-hypothesis sonar tracking \cite{Quach1}. A feature of MHT
algorithms is their ability to initiate tracks automatically in
clutter, as well as allowing for multiple targets. The article
\cite{Quach1} describes an adaptive re-initialisation procedure for
MHT. The state variable includes SNR, bearing, frequency and their
derivatives.  A track confidence test (based on $P_D$ and $P_{FA}$) is
employed to detect track loss, for instance, due to a manoeuvre. The
gate is then enlarged for a predetermined number of scans in order to
re-acquire the target.

\section{Discussion}\label{conc}
We have given an extensive review of the existing literature on
manoeuvring target models (section \ref{models}), manoeuvre detection
(section \ref{detec}), tracking of manoeuvring targets (section
\ref{filters}), and tracking of manoeuvring targets in clutter
(section \ref{clutter}).  Although the number of competing techniques
is large, the number of comparative performance evaluations is small.
A recent initiative \cite{Blair5} has been the definition of a
benchmark manoeuvring target tracking problem for phased array radar.

Conventional manoeuvring target models we have covered include random
input models, unknown deterministic input models, switching systems
and Singer's correlated process noise model. Several extensions to
these models have been investigated that use different stochastic
representations or higher order approximations.  Manoeuvre detection
strategies are characterised by the selection of a test statistic
based on the filter residuals using data windowing or exponential
forgetting together with a likelihood ratio test. Several
generalisations of these tests are available but do not appear to have
found significant application.  The use of additional sensors such as
imaging sensors is a promising means of enhancing the manoeuvre
detection capabilities of short-range radar tracking systems.

If false alarms are to be handled by a nearest-neighbour approach or
other correlation test, the accepted techniques for manoeuvring target
tracking can be classed as multi-level process noise, input
estimation, variable-dimension filtering, two-stage Kalman filtering,
and sub-optimal Bayesian filtering for switching systems including the
IMM and GPB algorithms.
The problem of tracking a manoeuvring target in clutter has been
formulated within the framework of Bayesian probability theory
using (i) a finite state Markov chain manoeuvre input and (ii)
a Markov chain switching model. The optimal Bayesian filter for
both these cases has combinatorially increasing complexity and therefore
$N$-scan truncation is generally applied to simplify the filter implementation.
The probabilistic data association filter, a single-scan Bayesian algorithm,
can be employed to extend the variable dimension filter and IMM filter
for tracking in clutter. 
The use of non-kinematic information such as measurement SNR,
when this is reliable, is a promising avenue for measurement
discrimination in clutter.

More recent approaches to the manoeuvring target tracking problem
include the expectation maximisation algorithm and MHT framework and
these deserve further comment.  The EM technique, which is a batch
algorithm, does not seem to be practicable in the cluttered case
without substantial modification to avoid enumeration of joint
manoeuvre and measurement associations.  It is however possible to
apply ideas from the recent probabilistic multi-hypothesis tracking
(PMHT) algorithm \cite{Streit4} to simplify these computations, thus
obtaining a practicable sub-optimal algorithm; the approaches
presented in \cite{Willett} and \cite{Logothetis2} are examples of
this. The key simplifying assumption in applying the EM algorithm in
this case is the supposition that the target may be associated with
all measurements in a scan simultaneously \cite{Pulford15}. This is
exploited via mixing probabilities to condense an $N$-dimensional sum
over all measurement-to-target associations over $N$ scans to a single
sum.  This assumption has no physical basis and also modifies the
tracking problem in a fundamental manner by assuming away the
combinatorial explosion of association hypotheses.  Numerical
simulations of PHMT-based EM algorithms in \cite{Willett} indicate
some improvement over the IMM-PDA algorithm (section \ref{immpda}) in
heavy clutter, although no comparisons with high performance
techniques such as MHT or GPB approaches appear to have been effected
as yet. Such comparisons would need to allow for the advantage of
smoothing over filtering, as well as the operational consequences of
increasing the latency in displaying tracks to the operator.

\subsection*{Acknowledgements}
This work was funded jointly by the Cooperative Research Centre for
Sensor Signal and Information Processing (CSSIP) and Defence Science
and Technology Organisation Australia under DSTO contract number
476189 (Viterbi Data Association Tracker for Over-the-Horizon Radar) \cite{Pulford11}.
The author is grateful to Dr Barbara La Scala for assisting in preparing
the literature review for the section on manoeuvre detection. 

\newpage
{\small
\bibliography{/rg/cssip/a/gwp/latex/refs-abc,manrefs}
\bibliographystyle{plain}
}
\end{document}